\documentclass{article}

\usepackage{amsmath,amsfonts,bm}

\def\eqref#1{equation~\ref{#1}}

\def\1{\bm{1}}

\DeclareMathAlphabet{\mathsfit}{\encodingdefault}{\sfdefault}{m}{sl}
\SetMathAlphabet{\mathsfit}{bold}{\encodingdefault}{\sfdefault}{bx}{n}

\def\gI{{\mathcal{I}}}

\usepackage[preprint]{neurips_2025}
\usepackage[utf8]{inputenc} 
\usepackage[T1]{fontenc}    
\usepackage{hyperref}       
\usepackage{url}            
\usepackage{booktabs}       
\usepackage{amsfonts}       
\usepackage{nicefrac}       
\usepackage{microtype}      
\usepackage{tikz}
\usepackage{forest}
\usepackage[normalem]{ulem} 
\usepackage{pifont}
\usepackage{enumitem}
\usepackage{float}
\usepackage{xcolor}
\useunder{\uline}{\ul}{}
\usetikzlibrary{shadows.blur,arrows.meta}
\usepackage{makecell} 
\usepackage{caption}
\usepackage{transparent}
\usepackage{tikz}
\definecolor{lightgray}{gray}{0.95}
\definecolor{good}{HTML}{FBDDE3}
\definecolor{bad}{HTML}{C9E9F7}
\definecolor{star}{HTML}{70AD47}

\usepackage{colortbl}

\newcommand{\shadeRow}[2]{%
    \rowcolor{white!#1!#2}
}

\title{PFMBench: Protein Foundation Model Benchmark}

\author{\textbf{Zhangyang Gao}\textsuperscript{1,2, ${*}$}, \textbf{Hao Wang} \textsuperscript{1, ${*}$}, \textbf{Cheng Tan} \textsuperscript{1,2,${*}$}, \textbf{Chenrui Xu} \textsuperscript{1}, \\
\textbf{Mengdi Liu} \textsuperscript{2,3}, \textbf{Bozhen Hu} \textsuperscript{1,2}, \textbf{Linlin Chao} \textsuperscript{1}, \textbf{Xiaoming Zhang} \textsuperscript{1, ${\dagger}$}, \textbf{Stan Z. Li} \textsuperscript{1,2, ${\dagger}$} 
\\
\textsuperscript{1} BioMap Research \\
\textsuperscript{2} AI Lab, Research Center for Industries of the Future, Westlake University \\
\textsuperscript{3} University of Chinese Academy of Sciences, China\\
\thanks{$^{\dagger}$ Corresponding Author, $^{*}$ Equal Contribution. Work done during internship at BioMap.}
}
\makeatletter
\def\thanks#1{\protected@xdef\@thanks{\@thanks
\protect\footnotetext{#1}}}
\makeatother

\begin{document}

\maketitle

\begin{figure}[h]
    \vspace{-13mm}
    \centering
    \includegraphics[width=4in]{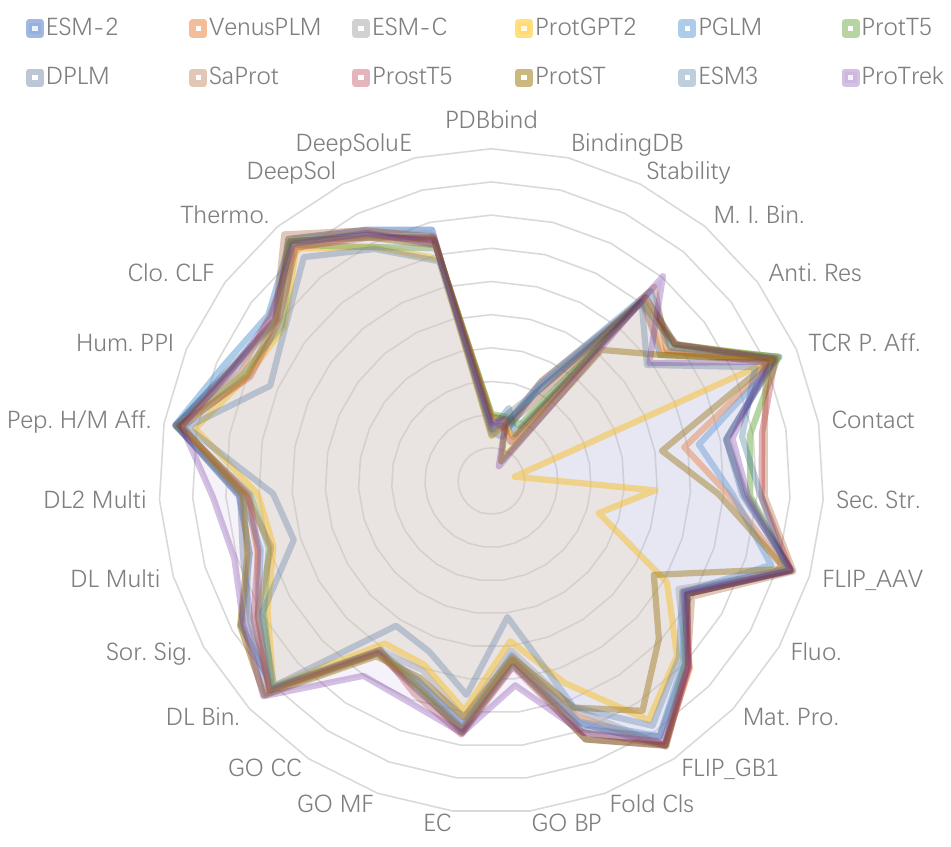}
    \caption{Protein foundation model performance on predictive tasks.} 
    \label{fig:task_relations}
\end{figure}

\begin{abstract}
    This study investigates the current landscape and future directions of protein foundation model research.   While recent advancements have transformed protein science and engineering, the field lacks a comprehensive benchmark for fair evaluation and in-depth understanding. Since ESM-1B, numerous protein foundation models have emerged, each with unique datasets and methodologies. However, evaluations often focus on limited tasks tailored to specific models, hindering insights into broader generalization and limitations. Specifically, researchers struggle to understand the relationships between tasks, assess how well current models perform across them, and determine the criteria in developing new foundation models.  To fill this gap, we present PFMBench, a comprehensive benchmark evaluating protein foundation models across 38 tasks spanning 8 key areas of protein science. Through hundreds of experiments on 17 state-of-the-art models across 38 tasks, PFMBench reveals the inherent correlations between tasks, identifies top-performing models, and provides a streamlined evaluation protocol. Code is available at \href{https://github.com/biomap-research/PFMBench}{\textcolor{blue}{GitHub}}.
\end{abstract}

\section{Introduction}


\vspace{-3mm}
Protein foundation models (PFMs) have garnered significant attention in recent years for their transformative potential in protein science and engineering. By training on large-scale protein datasets, these models capture intricate relationships between sequences, structures, and functions. Since the debut of ESM-1B \cite{rives2021biological} in 2021, a diverse array of PFMs—spanning various architectures and training paradigms—has emerged \cite{rives2021biological, lin2023evolutionary, hayes2025simulating, elnaggar2021prottrans, madani2023large, ferruz2022protgpt2, tan2025venusfactory, zhou2025protclip, elnaggar2023ankh, chen2024xtrimopglm, wang2024diffusion, susaprot, su2024protrek, xu2023protst, bjerregaard2025foundation, guo2025foundation,li2024progress}. Despite this rapid progress, prior models like ESM2 \cite{lin2023evolutionary} still dominate many bioengineering applications. This raises several pressing questions: Has the field reached a plateau and what is the next frontier for PFMs? Thus, a comprehensive and systematic benchmark is urgently needed.

\vspace{-3mm}
\begin{figure}[h]
    \centering
    \includegraphics[width=5.5in]{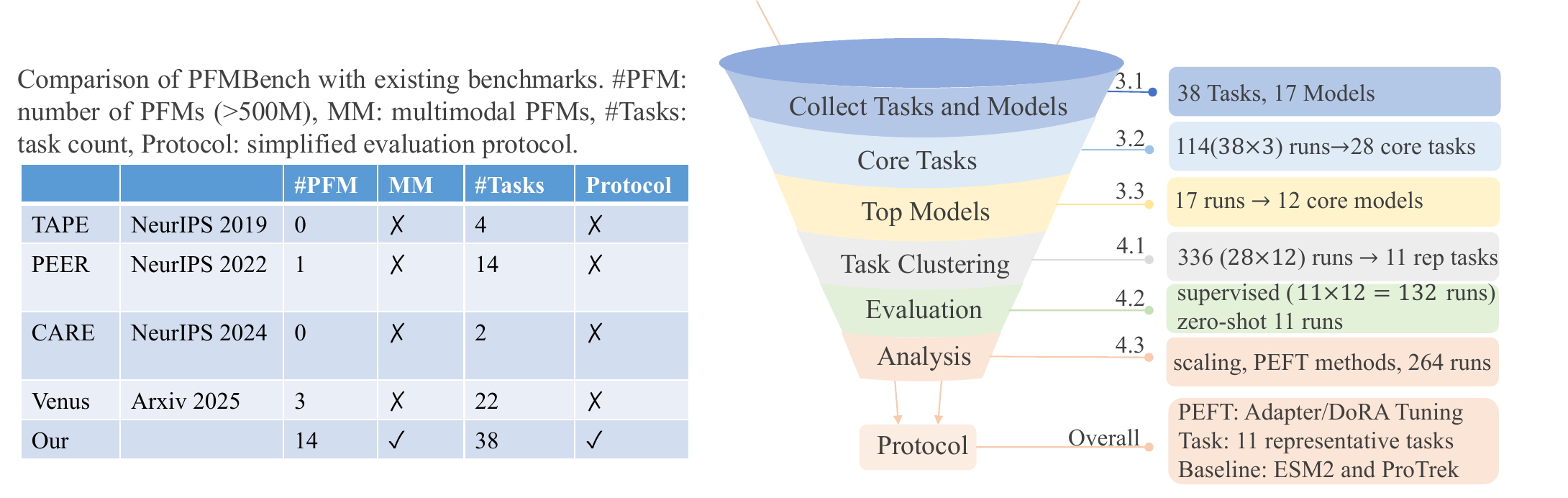}
    \vspace{-3mm}
    \caption{PFMBench: More tasks, multimodal PFMs, a simplified protocol, and hierarchical analysis.}
    \label{fig:cover}
\end{figure}
\vspace{-3mm}

Previous benchmarking efforts for protein models have either covered a limited set of tasks or were not explicitly designed for evaluating foundation models, as shown in Fig.~\ref{fig:cover}. In the context of protein foundation models (PFMs)—typically defined as models with at least 500 million parameters—most existing benchmarks fall short of providing comprehensive evaluation. For example, TAPE \cite{rao2019evaluating} assessed architectures such as Transformers \cite{vaswani2017attention}, LSTMs \cite{hochreiter1997long}, and ResNets \cite{he2016deep} across four tasks, but did not include any large-scale PFMs. PEER \cite{xu2022peer} evaluated models on 14 tasks but was limited to sequence-based architectures, with only ESM-1B \cite{rives2021biological} exceeding the 500-million-parameter threshold. CARE \cite{yang2024care} focused narrowly on two enzyme-related tasks: classification and retrieval.  More recently, VenusFactory \cite{tan2025venusfactory} introduced a unified benchmark spanning 22 tasks across five functional categories. However, it reported results for only three large sequence-based models, such as ESM2 \cite{lin2023evolutionary}, Ankh \cite{elnaggar2023ankh}, and ProtT5 \cite{elnaggar2021prottrans}, limiting its ability to capture the full spectrum of modern PFMs.

Multimodal PFMs are understudied in existing benchmarks, despite the field's rapid shift toward models that integrate sequence, structure, and functional data. For example, ESM3 \cite{hayes2025simulating}, GearNet \cite{zhangprotein}, and SaProt \cite{susaprot} have demonstrated strong performance on specialized tasks such as protein design and function prediction. However, their evaluations are often limited in scope, focusing on specific tasks or datasets, which impedes a systematic understanding of their limitations, generalizability, and cross-task performance. For instance, while ESM3 excels in protein design, its ability to generalize to other tasks remains largely unexplored. Similarly, GearNet and SaProt have shown promise in certain tasks, but their performance across broader protein function landscapes has yet to be thoroughly assessed. Consequently, it remains unclear under what conditions and how multimodal PFMs contribute to improved generalization capabilities.


A benchmark should not merely serve as a collection of tasks and models—it should also provide a streamlined protocol for model development. As both tasks and models become increasingly complex, exhaustively evaluating all models across all tasks becomes impractical and often fails to yield actionable insights. A more effective approach is to uncover the underlying relationships between tasks, identify a representative subset of tasks, and select a diverse yet informative set of models for focused evaluation. This strategy enables the benchmark to help researchers identify top-performing models for specific tasks and guide the development of new models—serving as a blueprint for future model evaluation, selection, and design.

To address this gap, we introduce PFMBench—a unified and comprehensive benchmark suite for protein foundation models. PFMBench spans 38 tasks across 8 categories, encompassing 19 sequence-based, sequence-structure, sequence-function, and multimodal PFMs. Both datasets and models are carefully curated to ensure robust, fair and meaningful comparisons. Through extensive evaluation, PFMBench offers detailed insights into the strengths and limitations of modern PFMs, and provide a simplified and useful protocol for future PFM development.

\section{Related Work}
\paragraph{Protein Foundation Models.} Protein foundation models (PFMs) have witnessed exponential growth in recent years, revolutionizing computational biology through self-supervised learning on vast protein sequence datasets. ESM-1b \cite{rives2021biological} pioneered large-scale protein modeling with a 650M parameter transformer trained on 65 million protein sequences via masked language modeling. This trajectory continued with ESM-2 and ESMC models \cite{lin2023evolutionary}, which demonstrated enhanced representation learning for protein structure and function through refined architecture and expanded training data. The ESM family evolved further with ESM3 \cite{hayes2025simulating}, scaling to 98B parameters and incorporating structure-aware training to achieve state-of-the-art performance on zero-shot fitness prediction and structure modeling.  ProtT5 \cite{elnaggar2021prottrans} adapted the T5 architecture to proteins, scaling to 3B and 11B parameters with span-masking objectives, establishing strong baselines for protein sequence-to-sequence tasks. The generative approach was pioneered by ProGen \cite{madani2023large}, a 1.2B parameter conditional generation model, and ProtGPT2 \cite{ferruz2022protgpt2}, a 738M parameter GPT-2-based model for de novo protein sequence generation. VenusPLM \cite{tan2025venusfactory} employed transformer-based architectures with modular fine-tuning capabilities for enzyme engineering and protein function prediction. Multimodal approaches emerged with ProtCLIP \cite{zhou2025protclip}, aligning protein sequences with biological text through function-informed pre-training. ANKH \cite{elnaggar2023ankh} built upon ProtT5's architecture to optimize data efficiency through systematic ablation studies. xTrimoPGLM \cite{chen2024xtrimopglm} adpot GLM's training paradigm to protein sequences, expanding the model size to 100B. Other significant contributions include DPLM \cite{wang2024diffusion}, leveraging deep learning for protein language modeling; SaProt \cite{susaprot}, focusing on structure-aware protein representation learning; ProtRek \cite{su2024protrek}, specialized in protein sequence retrieval and knowledge integration; and ProST \cite{xu2023protst}, which incorporates biomedical texts to guide protein function learning. Together, these diverse foundation models have transformed protein research by enabling unprecedented advances in structure prediction, functional annotation, and protein design through their ability to learn complex evolutionary and structural patterns from sequence data.

\paragraph{Protein Benchmarks.} Protein foundation model benchmarks have evolved significantly, transitioning from early efforts like TAPE \cite{rao2019evaluating}, which evaluated small models on a limited set of tasks, to more comprehensive frameworks. PEER \cite{xu2022peer} expanded the scope by introducing a multi-task benchmark encompassing diverse protein understanding tasks, including function prediction and protein-protein interactions. BeProf \cite{wang2024comprehensive} further contributed by evaluating deep learning-based protein function prediction models in different application scenarios. Recent benchmarks like VenusFactory \cite{tan2025venusfactory} have integrated a broader range of pre-trained models and datasets, yet they often lack consideration for multimodal approaches. Beyond predictive benchmarks, initiatives like ProteinGym \cite{notin2023proteingym}, ProteinInvBench \cite{gao2023proteininvbench} and ProteinBench \cite{ye2024proteinbench} have introduced frameworks for evaluating protein mutation effects, inverse folding and protein design, respectively. These benchmarks have progressively incorporated more diverse tasks, models—including large pre-trained language models and multimodal approaches—and sophisticated evaluation metrics, thereby playing a crucial role in tracking progress, identifying state-of-the-art methods, and guiding future research. However, current benchmarks do not foucus on protein foundation models, especially multimodal foundation models, also do not provide a streamlined evaluation protocol for these models.

\paragraph{Parameter-Efficient Fine-Tuning.} Recent advances in parameter-efficient fine-tuning (PEFT) have enabled the adaptation of large pre-trained models by updating only a small subset of their parameters. Adapter-based methods insert trainable modules between frozen layers \cite{houlsby2019parameter, pfeiffer2020adapterfusion}, while Low-Rank Adaptation (LoRA) approximates weight updates using low-rank matrices \cite{hu2022lora}. Prompt-based techniques—such as prefix tuning \cite{li2021prefix} and prompt tuning \cite{lester2021power}—optimize soft prompts within the input embeddings, avoiding changes to the model weights. Other approaches, including BitFit (which updates only bias terms) \cite{zaken2022bitfit}, IA3 (which scales intermediate activations) \cite{liu2022few}, and QLoRA (which enables quantized fine-tuning) \cite{dettmers2023qlora}, further improve efficiency. Hybrid strategies that combine multiple techniques have also emerged \cite{hetowards}. Recent innovations include AdaLoRA, which dynamically adjusts rank allocation during training \cite{zhang2023adalora}; MoeLoRA, which integrates mixture-of-experts into LoRA for enhanced scalability \cite{wu2024mixture}; DoRA, which decomposes weights into magnitude and direction for targeted adaptation \cite{mao2024dora}; and LoCA, which introduces location-aware cosine adaptation for more precise updates \cite{du2025loca}. Collectively, these developments continue to improve the efficiency, flexibility, and effectiveness of PEFT for large language models. This research select Adapter, LoRA, AdaLoRA, DoRA and IA3  as the representative methods for performance comparison.

\clearpage
\section{Method}
\subsection{PFMBench Framework}
\begin{figure}[h]
    \centering
    \includegraphics[width=5in]{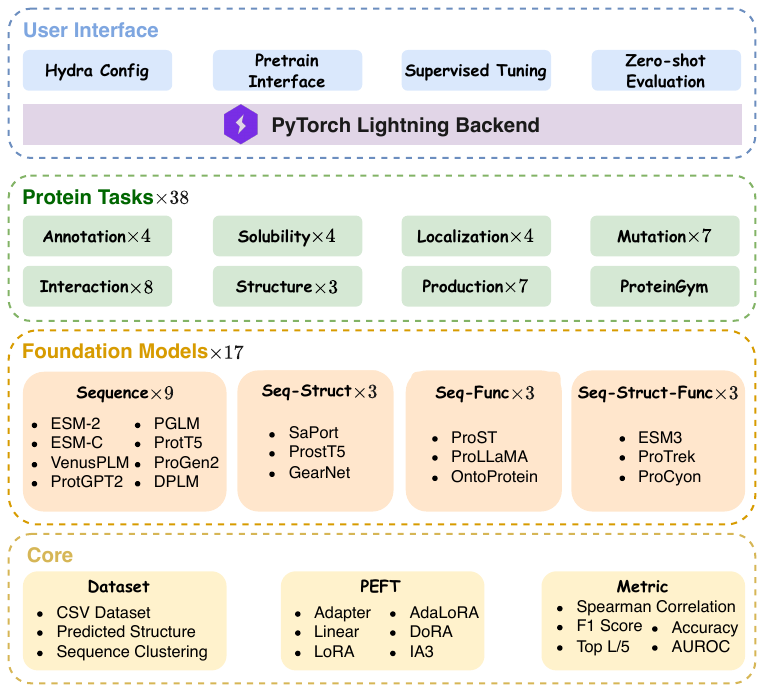}
    \caption{The Overall framework of PFMBench.  The framework includes: (1) a user-friendly interface, (2) enumerious downstream tasks, and (3) a comprehensive set of foundation models. Diverse datasets, parameter-efficient tuning methods, and evaluation metrics are integrated. The modular design allows users to easily swap components, customize models, tasks and metrics.}  
    \label{fig:framework}
\end{figure}

\paragraph{Framework.}
As shown in Figure \ref{fig:framework}, PFMBench comprises three main components: (1) a user-friendly interface, (2) a suite of downstream tasks, and (3) a comprehensive collection of foundation models. Designed with modularity in mind, the framework allows users to swap components and customize the evaluation process with ease. We employ Hydra to parse configuration files and PyTorch Lightning to manage model fine-tuning. To our knowledge, PFMBench is the largest and most comprehensive benchmark for protein foundation models, covering 38 tasks across 17 models. 

\paragraph{Data Contribution.}
For each dataset, we retrieve protein structures from the AF2DB \cite{varadi2022alphafold} when available; otherwise, we use ESMFold \cite{lin2023evolutionary} to generate the rank-1 protein structure. To standardize evaluation, we enforce a 30\% sequence similarity cutoff when splitting data, resulting in an 8:1:1 ratio for training, validation, and test sets. Mutation datasets are exempt from this splitting due to their high similarity to wild-type sequences; thus, we retain their original train/validation/test partitions.

\paragraph{Protocol Contribution.} 
Evaluating all models and tasks is impractical, especially when aiming to provide guidance for developing new foundation models. We believe that simplifying the selection of tasks and models is equally important, as it highlights the key insights. Through hundreds of experiments, we provide a hierarchical analysis that results in a streamlined protocol: (1) Baseline: select either the sequence-only ESM2 or the multimodal ProTrek; (2) Task: filter 11 representative tasks from the original 38 tasks; (3) PEFT: adopt either the transformer-adapter or the DoRA tuning.

\vspace{-3mm}

\subsection{Supported Tasks}

\paragraph{Core Tasks.} PFMBench includes 38 tasks spanning diverse domains, covering both supervised and zero-shot learning. Supervised tasks are grouped into seven categories: Annotation, Solubility, Localization, Mutation, Interaction, Structure, and Production. Definitions, metrics, and impacts for each category are detailed in Appendix~\ref{sec:appendix_tasks}. Datasets are split into training, validation, and test sets using an 8:1:1 ratio with a 30\% sequence similarity threshold, except for mutation datasets. We evaluate ESM2-Adapter on all tasks, averaging results over three runs, with bias calculated as the absolute difference between the best and worst runs divided by the average performance (see Table~\ref{tab:task_lists}). To ensure unbiased evaluation, we designates 28 tasks with a bias below 5\% as core tasks. 

\vspace{-3mm}

\begin{table}[h]
    \caption{PFMBench Tasks span eight categories, detailing training, validation, and test sample counts per task with references. Symbols $\triangle$ and \ding{73} indicate datasets with sequence or sequence-structure pairs, as used in benchmarks like TAPE \cite{rao2019evaluating}, PEER \cite{xu2022peer}, Venus \cite{tan2025venusfactory}, and our framework. ESM2-Adapter's mean and bias performance are shown, with core tasks having bias below 5\%.}
    \label{tab:task_lists}
    \resizebox{1.0 \columnwidth}{!}{\begin{tabular}{lllllllllllll}
    \toprule
    Task                  & Metric                & Train & Val & Test  & TAPE  & Peer       & Venus               & Our                        & Mean       & Bias(\%)  & Core                 \\ \midrule
    \multicolumn{8}{l}{\textbf{Annotation}}                                                                                                                                                        \\
    Cellular Component   \cite{ashburner2000gene}      &  F1 Score          & 11196 & 1398 & 1400   &     &       & \ding{73} & \ding{73} & 0.6130 & 0.26\%    & \checkmark \\
    Molecular Function  \cite{ashburner2000gene}   &   F1 Score             & 22291 & 2785 & 2787   &     &       & \ding{73} & \ding{73} & 0.6488 & 0.38\%    &  \checkmark                         \\
    Biological Process  \cite{ashburner2000gene}     &    F1 Score           & 21395 & 2662 & 2664   &    &        & \ding{73} & \ding{73} & 0.5412     & 0.79\%    &  \checkmark                         \\
    Enzyme Commission \cite{bairoch2000enzyme}    &    F1 Score                 & 13090 & 1465 & 1604   &     &       & \ding{73} & \ding{73} & 0.7379 & 0.09\%    & \checkmark \\ \midrule
    \multicolumn{8}{l}{\textbf{Solubility}}                                                                                                                                                        \\
    DeepSol \cite{khurana2018deepsol}           & AUROC                   & 55465 & 6932 & 6934 &  & $\triangle$ & \ding{73} & \ding{73} & 0.8467 & 0.23\%    & \checkmark \\
    DeepSoluE \cite{wang2023prediction}        & AUROC                  & 11627 & 1452 & 1454   &       &     & \ding{73} & \ding{73} & 0.7699     & 1.10\%    &  \checkmark                         \\
    ProtSolM \cite{tan2024protsolm}             & AUROC                & 57378 & 7171 & 7173   &     &       & \ding{73} & \ding{73} & 0.8572 & 0.93\%    & \checkmark \\
    eSOL  \cite{chen2021structure}            & Spearman                    & 2481 & 309 & 311      &    &        & \ding{73} & \ding{73} & 0.2761 & 38.3\%   &                           \\ \midrule
    \multicolumn{8}{l}{\textbf{Localization}}                                                                                                                                                      \\
    DeepLoc Multi \cite{almagro2017deeploc}     & Accuracy                    & 6992 & 749 & 751  &    & $\triangle$ & \ding{73} & \ding{73} & 0.7666 & 1.27\%    &     \checkmark                      \\
    DeepLoc2 Multi  \cite{thumuluri2022deeploc}   &  F1 Score                    & 21949 & 2743 & 2744 &  & $\triangle$ & \ding{73} & \ding{73} & 0.7505     & 0.16\%    & \checkmark \\
    DeepLoc Binary \cite{almagro2017deeploc}      & AUROC                   & 6887 & 846 & 848      &    &        &                            & \ding{73} & 0.9338 & 0.42\%    &        \checkmark                   \\
    Sorting Signal \cite{thumuluri2022deeploc}   &   F1 Score                   & 1484 & 185 & 186      &    &        & $\triangle$                 & \ding{73} & 0.8598 & 0.24\%    & \checkmark \\ \midrule
    \multicolumn{8}{l}{\textbf{Mutation}}                                                                                                                                                          \\
    PETA\_CHS\_Sol  \cite{tan2024peta}       & Spearman                & 3872 & 484 & 484      &      &      & $\triangle$                 & \ding{73} & 0.2738     & 12.5\% &                           \\
    PETA\_LGK\_Sol    \cite{tan2024peta}      & Spearman               & 15308 & 1914 & 1914   &      &      & $\triangle$                 & \ding{73} & 0.1558     & 21.7\% &                           \\
    PETA\_TEM\_Sol  \cite{tan2024peta}      & Spearman                 & 6444 & 808 & 808      &      &      & $\triangle$                 & \ding{73} & 0.1433 & 27.0\% &                           \\
    FLIP\_AAV  \cite{dallago2flip}   & Spearman      & 66066 & 16517 & 16517 &     &       & $\triangle$                 & \ding{73} & 0.9412 & 0.13\%    &           \checkmark                \\
    FLIP\_GB1   \cite{dallago2flip}  & Spearman     & 6988 & 1745 & 1745    &       &     & $\triangle$                 & \ding{73} & 0.9517 & 0.13\%    & \checkmark \\
    TAPE\_Stability \cite{rao2019evaluating}       & Spearman                & 55182 & 6897 & 6898 & $\triangle$  & $\triangle$ & $\triangle$                 & \ding{73} & 0.3211     & 4.01\%    & \checkmark \\
    TAPE\_Fluorescence  \cite{rao2019evaluating}       & Spearman            & 21446 & 5362 & 27217 & $\triangle$  & $\triangle$ & $\triangle$                 & \ding{73} & 0.6812 & 0.21\%   &  \checkmark                         \\
    $\beta$-lactamase activity \cite{gray2018quantitative} & Spearman  & 4158 & 520 & 520    &  & $\triangle$ &                            & \ding{73} &  0.5740           & 21.6\%       &                           \\ \midrule
    \multicolumn{8}{l}{\textbf{Interaction}}                                                                                                                                                       \\
    Human-PPI  \cite{pan2010large}      & AUROC                      & 30133 & 270 & 195 &    & $\triangle$ &                            & \ding{73} & 0.4828     & 0.00\%    & \checkmark \\
    Yeast-PPI   \cite{guo2008using}       & AUROC                    & 4157 & 83 & 335   &    & $\triangle$ &                            & \ding{73} & 0.5343     & 12.8\%   &                           \\
    PPI affinity \cite{moal2012skempi}    & Spearman           & 2421 & 203 & 326  &    & $\triangle$ &                            &        \ding{73}                    & -0.0047 & 114.3\% &                           \\
    PDBbind   \cite{liu2017forging}      & Spearman                      &   14687  & 1835 & 1836  & & $\triangle$ &                            & \ding{73}  &  0.1677    &    4.14\%      & \checkmark                 \\
    BindingDB  \cite{liu2007bindingdb}     & Spearman                      & 9039 & 1115 & 1139  &   & $\triangle$ &                            & \ding{73} & 0.1922 & 3.02\%    &    \checkmark                       \\
    Metal ion Binding  \cite{hu2022exploring}    & Accuracy               & 5740 & 717 & 718      &     &       & \ding{73} & \ding{73} & 0.7066     & 2.43\%    &           \checkmark                \\
    Pept.HLA/MHC Aff. \cite{wu2023ccbhla}     & AUROC         & 57357 & 7008 & 8406   &            &         &                   & \ding{73} & 0.9631     & 0.00\%    & \checkmark \\
    TCR PMHC Affinity  \cite{koyama2023attention}     & AUROC               & 19264 & 2265 & 2482   &            &      &                      & \ding{73} & 0.9312     & 0.00\%    &          \checkmark                 \\ \midrule
    \multicolumn{8}{l}{\textbf{Structure}}                                                                                                                                                         \\
    Contact prediction \cite{yang2020improved}   & Top L/5                 & 12005 & 1500 & 1501  & $\triangle$  & $\triangle$ &                            & \ding{73} &    0.7199        &    0.40\%       &    \checkmark                       \\
    Fold classification \cite{lo2000scop}     & Accuracy         & 13034 & 1628 & 1630 &     & $\triangle$ &                            & \ding{73} & 0.7859 & 0.31\%    &       \checkmark                    \\
    Secondary structure \cite{klausen2019netsurfp}  & Accuracy     & 67007 & 8365 & 8262  & $\triangle$ & $\triangle$ &                            & \ding{73} & 0.7601     & 0.00\%    & \checkmark \\ \midrule
    \multicolumn{8}{l}{\textbf{Production}}                                                                                                                                                        \\
    Optimal PH  \cite{gado2023deep}      & Spearman                     & 7669 & 958 & 959      &     &        &                            & \ding{73} & 0.0564     & 17.6\%   &                           \\
    DeepET\_Topt   \cite{li2022learning}     & Spearman                   & 1479 & 184 & 185      &    &         & \ding{73} & \ding{73} & 0.2628     & 7.00\%    & \\
    Cloning CLF  \cite{wang2014predppcrys}       & AUROC                  & 22223 & 2777 & 2778   &            &        &                     & \ding{73} & 0.8160 & 0.51\%    & \checkmark \\
    Material Production  \cite{wang2014predppcrys}    & Accuracy              & 22196 & 2773 & 2775   &            &          &                   & \ding{73} & 0.7982     & 0.00\%    & \checkmark \\
    Enzyme Eff.  \cite{li2022deep}  & Spearman        & 10363 & 1298 & 1290   &            &      &                       & \ding{73} & 0.2173     & 58.2\%   &                           \\
    Antib. Res.  \cite{hu2022exploring}    & Accuracy           & 2703 & 336 & 339      &            &         &                    & \ding{73} & 0.6185     & 2.23\%    &           \checkmark                \\
    Thermostability  \cite{jarzab2020meltome}    & AUROC                  & 33474 & 4184 & 4184 &    &         & \ding{73} & \ding{73} &    0.9553        &    1.27\%       &     \checkmark                       \\ \midrule
    \multicolumn{8}{l}{\textbf{Zero-shot}}                                                                                                                                                         \\
    ProteinGym  \cite{notin2023proteingym}    & Spearman          &  &      &          &                  &            & \ding{73} & \ding{73} &    0.4390        &    0\%       &   \checkmark    \\ \bottomrule       
    \end{tabular}}
\end{table}
\vspace{-3mm}

\subsection{Supported Models}
\paragraph{Core Models.} PFMBench supports a broad spectrum of protein foundation models, as summarized in Table \ref{tab:models}. To ensure a fair comparison, we select models with parameter counts close to 1B when multiple versions are available. Based on input data modalities, the models are categorized into four groups: (1) sequence-only models, (2) sequence-structure models, (3) sequence-function models, and (4) sequence-structure-function models. To establish a consistent evaluation baseline, we assess all models on the enzyme commission (EC) classification task under the adapter tuning setting. Models that achieve at least 85\% of ESM2’s performance are selected as core models for further evaluation.

\vspace{-3mm}
\begin{table}[h]
    \centering
    \caption{Models in PFMBench. The table lists the models, architecture types, number of parameters, publication states, code sources. We report the Enzyme Commission (EC) results.}
    \label{tab:models}
    \resizebox{0.8 \columnwidth}{!}{\begin{tabular}{lllllll} \toprule
    Model            & Core                & Architecture       & \# Params & Publication                      & EC  & Code \\ \midrule
    \multicolumn{6}{l}{\textbf{Sequence}}                                                                                                 \\
    ESM-2 \cite{lin2023evolutionary}           & \checkmark & Encoder            & 650M      & Science 23               & 0.7358     &  \href{https://huggingface.co/facebook/esm2_t33_650M_UR50D}{HF}     \\
    VenusPLM \cite{tan2025venusfactory}        & \checkmark & Encoder            & 300M      & Arxiv 25                 & 0.7519   & \href{https://huggingface.co/AI4Protein/VenusPLM-300M}{HF}        \\
    ESM-C            & \checkmark & Encoder            & 600M      & Blog 25                  & 0.7169     & \href{https://huggingface.co/EvolutionaryScale/esmc-600m-2024-12}{HF}      \\
    ProtGPT2 \cite{ferruz2022protgpt2}        &    \checkmark                       & Decoder            & 738M      & Nat. Commun. 22          & 0.6969       & \href{https://huggingface.co/nferruz/ProtGPT2}{HF}    \\
    ProGen2  \cite{nijkamp2023progen2}        &                           & Decoder            & 764M      & Cell Syst. 23            & 0.6198    & \href{https://github.com/enijkamp/progen2}{GitHub}       \\
    xTrimoPGLM \cite{chen2024xtrimopglm}   & \checkmark & Encoder-Decoder    & 1B        & Nat. Methods 25          & 0.7466      & \href{https://huggingface.co/biomap-research/proteinglm-1b-mlm}{HF}     \\
    ProtT5  \cite{elnaggar2021prottrans}         & \checkmark & Encoder-Decoder    & 3B        & TPAMI 21                 & 0.7620 &     \href{https://huggingface.co/Rostlab/prot_t5_xl_uniref50}{HF}     \\
    DPLM    \cite{wang2024diffusion}         & \checkmark & Encoder+Diffusion & 650M      & ICLM 24                  & 0.7552     &  \href{https://github.com/bytedance/dplm}{GitHub}    \\ \midrule
    \multicolumn{6}{l}{\textbf{Sequence-Structure}}                                                                                       \\
    SaPort  \cite{susaprot}         & \checkmark & Encoder            & 650M      & ICLR 24                  & 0.7514     & \href{https://huggingface.co/westlake-repl/SaProt_650M_AF2}{HF}      \\
    ProstT5 \cite{heinzinger2024bilingual} & \checkmark & Encoder-Decoder    & 3B        & NAR Gen. Bio. 24 & 0.7683    & \href{https://github.com/mheinzinger/ProstT5}{GitHub}       \\
    GearNet  \cite{zhangprotein}        &                           & GNN                & 20M       & ICLR 23                  & 0.5860      & \href{https://github.com/DeepGraphLearning/GearNet}{GitHub}     \\ \midrule
    \multicolumn{6}{l}{\textbf{Sequence-Function}}                                                                                        \\
    ProtST \cite{xu2023protst}           & \checkmark & Encoder            & 750M      & ICML 23                  & 0.7176       & \href{https://github.com/DeepGraphLearning/ProtST}{GitHub}    \\
    ProLLaMA   \cite{lv2025prollama}      &                           & Decoder            & 6.7B      & IEEE TAI 25                 & 0.5475      & \href{https://github.com/PKU-YuanGroup/ProLLaMA?tab=readme-ov-file}{GitHub}     \\
    OntoProtein  \cite{zhangontoprotein}    &                           & Encoder            & 420M      & ICLR 22                  & 0.6287     & \href{https://github.com/zjunlp/OntoProtein}{GitHub}      \\ \midrule
    \multicolumn{6}{l}{\textbf{Sequence-Structure-Function}}                                                                              \\
    ProCyon  \cite{queen2024procyon}        &                           & Decoder            & 11B       & Arxiv 24                 & 0.1909    & \href{https://github.com/mims-harvard/ProCyon}{GitHub}       \\
    ESM3 \cite{hayes2025simulating}            & \checkmark & Encoder            & 1.4B      & Science 25               & 0.6483     & \href{https://github.com/evolutionaryscale/esm}{GitHub}       \\
    ProTrek \cite{su2024protrek}         & \checkmark & Encoder            & 650M      & Arxiv 24                 & 0.7641  & \href{https://github.com/westlake-repl/ProTrek}{GitHub}  \\ \bottomrule
    \end{tabular}}
\end{table}

\vspace{-3mm}
\subsection{Supported Tuning Methods}
PFMBench offers diverse parameter efficient fine-tuning (PEFT) methods: linear probing, adapter tuning, IA$^3$, LoRA, AdaLoRA, and DoRA, with a unified interface for seamless switching. 

\vspace{-1mm}
\paragraph{Adapter Tuning \& Linear Probing.}
We extract features using the pretrained model and employ a 6-layer transformer as a task-specific adapter with a hidden size of 480 and 20 attention heads. In Linear probing setting, we the transformer adapter is replaced with a linear layer. Without additional explanation, we report adapter tuning results in the main text. 

\vspace{-1mm}
\paragraph{Other Tuning Methods.} \textbf{LoRA} decomposes attention and feedforward layer weight updates into the product of two low-rank matrices, which are the only trainable components during finetuning~\cite{hu2022lora}. \textbf{IA$^3$} introduces trainable multiplicative scalars into the attention and MLP sublayers, modulating the flow of information through each component~\cite{liu2022few}. \textbf{AdaLoRA} dynamically adjusts rank allocation during training~\cite{mao2024dora}. \textbf{DoRA} decomposes weights into magnitude and direction for targeted adaptation~\cite{zhang2023adalora}. We implement these methods using the PEFT library~\cite{peft}.

\vspace{-1mm}
\paragraph{Hyper-parameters.} All models are trained for up to 50 epochs using AdamW with a batch size of 64 and early stopping after 5 patience epochs. Optimal learning rate is selected from \{1e-5, 1e-4\}.
\vspace{-3mm}
\section{Experiments}
\vspace{-2mm}
We conduct systematic experiments  to answer the following questions:
\begin{itemize}[leftmargin=1.5em]
    \item \textbf{Q1: Supervised Tuning.} How are different supervised downstream tasks correlated, and can a minimal, representative subset of tasks be identified to efficiently benchmark pre-trained models?
    \item \textbf{Q2: Zero-shot Evaluation.} Can zero-shot protocols reliably evaluate protein foundation models?
    \item \textbf{Q3: PEFT Strategies.} Which PEFT methods are more effective for protein tasks?
    \item \textbf{Q4: Scaling.} How does model performance improve with increased model size?
\end{itemize}

\subsection{Supervised Tunning (Q1)}
\paragraph{Task Correlations.} We evaluate the adapter tuning performance of 12 core models across 28 core tasks, with the complete results provided in the appendix (Table~\ref{tab:full_model_performance_adapter}) due to space constraints. We analyze task relationships using Spearman correlation and visualize the results in Figure~\ref{fig:task_relations}, where p-values greater than 0.05 are marked with \ding{55}. Finally, the 28 core tasks are grouped into 11 clusters based on their correlations, and the selected \textbf{representative tasks} (marked as \textcolor{star}{\ding{73}}).
\begin{figure}[h]
    \vspace{-4mm}
    \centering
    \includegraphics[width=4.5in]{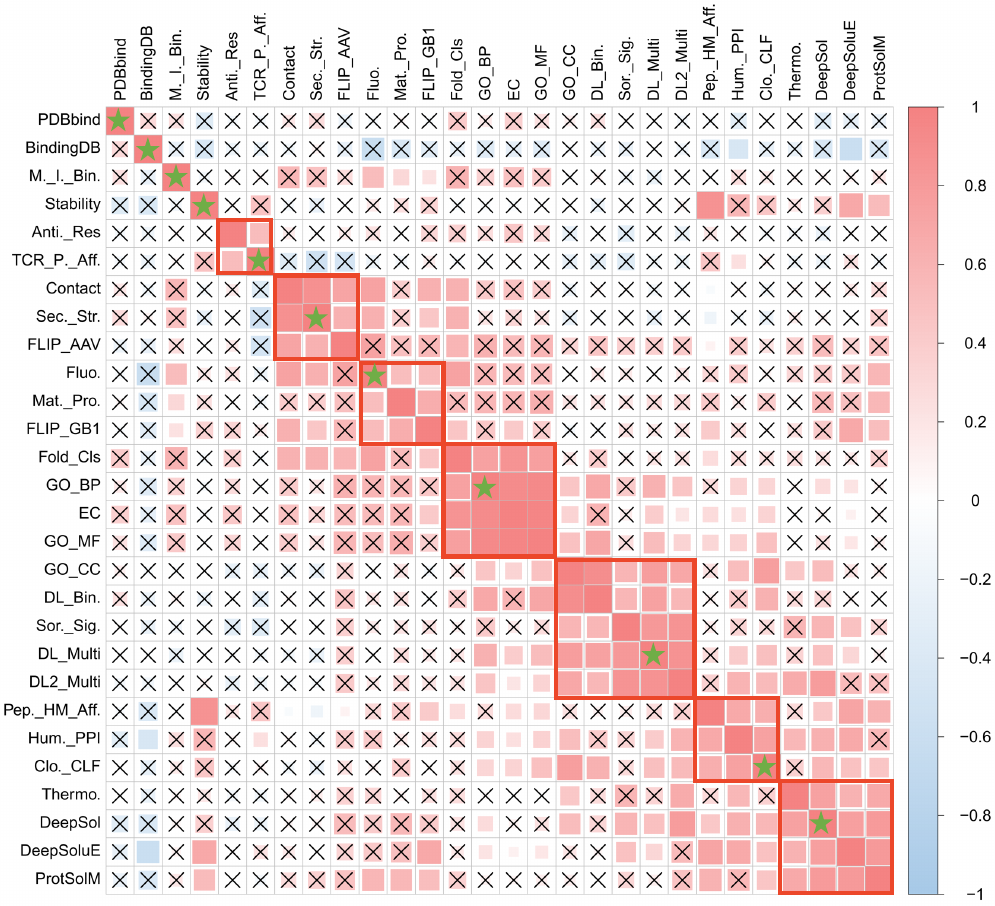}
    \caption{Task relations in supervised tuning.} 
    \label{fig:task_relations}
\end{figure}
\vspace{-4mm}

\paragraph{Core Model Performance on Representative Tasks.} Table~\ref{tab:core_model_performance} summarizes the performance of 12 core models on 11 representative tasks. Poorly performing tasks are excluded due to the challenges adapter tuning faces in learning them. Upon analyzing the poorly performing datasets, we observe that the newly implemented 30\% sequence identity split introduces significant challenges for model learning. While the stability performance under the original split aligns with SaProt \cite{susaprot}, the new split proves to be more demanding. Interaction tasks, requiring paired sequence embeddings processed via transformer adapters, remain particularly challenging, underscoring the need for PLMs tailored for interaction prediction, as current models are trained solely on single sequences.


\begin{table}[H]
    \vspace{-4mm}
    \caption{Core model results on representative tasks. \textbf{Best} and \underline{second-best} ones are highlighted. }
    \label{tab:core_model_performance}
    \resizebox{1.0 \columnwidth}{!}{\begin{tabular}{l
        >{\color{gray}}l
        >{\color{gray}}l
        >{\color{gray}}l
        lllllllllllll} \toprule
                     & PDBBind          & Bind. DB    & Stability      & Anti.Res.        & Mat.Pro.         & EC               & M. I. Bin.    & Sec. Str.        & DL2 M.    & Clo. CLF        & DeepSol    & \#Win      \\ \midrule
    \multicolumn{11}{l}{\textbf{Sequence}}  \\ 
    ESM-2  \cite{lin2023evolutionary}             & 0.14677          & 0.13692   & 0.32112       & 0.63422          & 0.81189          & 0.73578          & 0.71170          & 0.76375          & {\ul 0.76191}    & 0.80586          & \textbf{0.84494} & --  \\
    VenusPLM \cite{tan2025venusfactory}        & 0.16536          & 0.16834    & \textbf{0.33907}      & 0.64602          & \textbf{0.82018} & 0.75194          & 0.70195          & 0.71637          & 0.73814          & {\ul 0.83172}    & 0.82775    & 50\%  \\  \shadeRow{70}{bad}
    ESM-C            & 0.14692          & {\ul 0.20716}  & 0.29976  & 0.67257          & 0.81009          & 0.71694          & 0.70195          & 0.76777          & 0.75395          & 0.81033          & 0.84171    & 38\%    \\ \shadeRow{50}{bad}
    ProtGPT2 \cite{ferruz2022protgpt2}         & 0.13503          & 0.17169  & 0.14803        & 0.68437          & 0.76757          & 0.69687          & 0.71170          & 0.49371          & 0.70341          & 0.77730          & 0.78883     & 13\%  \\ 
    PGLM   \cite{chen2024xtrimopglm}   & 0.16877          & 0.16884   & {\ul 0.33127}       & 0.67257          & 0.79495          & 0.74659          & {\ul 0.74513}    & 0.72842          & 0.74772          & \textbf{0.83638} & 0.82160      & 50\%   \\
    ProtT5  \cite{elnaggar2021prottrans}         & \textbf{0.20105} & 0.19730    & 0.18638        & {\ul 0.68732}    & 0.80072          & 0.76201          & 0.72145          & 0.77978          & 0.72624          & 0.78485          & 0.78741   & 50\%     \\
    DPLM  \cite{wang2024diffusion}            & 0.13659          & 0.17408    & 0.29440        & {\ul 0.68732}    & 0.80144          & 0.75521          & 0.70056          & 0.75695          & 0.75759          & 0.81247          & 0.82841    & 38\%    \\ \midrule
    \multicolumn{11}{l}{\textbf{Sequence-Structure}}  \\ \shadeRow{70}{good}
    SaProt  \cite{susaprot}           & 0.15549          & 0.16557    & 0.24804       & 0.65782          & 0.81081          & 0.75144          & 0.71031          & \textbf{0.82389} & 0.74006          & 0.81206          & {\ul 0.84364}  & 50\%  \\ \shadeRow{70}{good}
    ProstT5 \cite{heinzinger2024bilingual} & 0.18344          & 0.16642     & 0.13032      & \textbf{0.69027} & {\ul 0.81622}    & \textbf{0.76829} & 0.72145          & {\ul 0.81397}    & 0.73190          & 0.79853          & 0.81937     & \underline{63\%}    \\ \midrule
    \multicolumn{11}{l}{\textbf{Sequence-Function}}  \\ \shadeRow{50}{bad}
    ProtST  \cite{xu2023protst}         & {\ul 0.19514}    & 0.18886     & 0.06623      & 0.63422          & 0.69261          & 0.71761          & 0.51532          & 0.68468          & 0.74886          & 0.80714          & 0.81951    & 13\%    \\ \midrule
    \multicolumn{11}{l}{\textbf{Sequence-Structure-Function}}  \\ \shadeRow{50}{bad}
    ESM3  \cite{hayes2025simulating}           & 0.15572          & \textbf{0.22519}   & 0.15650 & 0.58407          & 0.77514          & 0.64830          & 0.70334          & {0.81264}          & 0.65853          & 0.77391          & 0.78106      & 13\%   \\  \shadeRow{50}{good}
    ProTrek  \cite{su2024protrek}        & 0.17322          & 0.19230   & 0.04924       & 0.59292          & 0.81477          & {\ul 0.76408}    & \textbf{0.80362} & {0.77363}          & \textbf{0.83944} & 0.82612          & 0.83427 & \textbf{75\%} \\ \bottomrule       
    \end{tabular}}
\end{table}

\paragraph{Do existing PLMs truly outperform ESM2?} For the remaining 8 representative tasks, we compare each model against ESM2 and calculate the winning rate (\#Win), which is defined as the proportion of tasks where a model outperforms ESM2.  From the \#Win metric, we observe that: 
\vspace{-2mm}
\begin{itemize}[leftmargin=1.5em]
    \item \textbf{Sequence-based PLMs.} All sequence-based PLMs achieve no more than a 50\% winning rate against ESM2, indicating that they could not outperform ESM2 on the representative tasks.
    \item \textbf{Decoder-only Model.} The decoder-only model ProtGPT2 performs the worst on these tasks, with a winning rate of only 13\% on representative tasks. This suggests that the decoder-only architecture is currently unsuitable for protein understanding.
    \item \textbf{Multimodal PLMs.} Multimodal PLMs achieve the highest winning rates, with ProTrek attaining a 75\% winning rate on representative tasks. This success is attributed to the effective semantic alignment of sequence and function information during the pre-training stage.
    \item \textbf{Challenges with Function Data.} ESM3 and ProtST show low winning rates (13\%) due to noisy or insufficient function data, emphasizing the need for high-quality, large-scale datasets. For example, ProTrek excels when trained on such cleaned, large-scale annotations.
\end{itemize}



\vspace{-3mm}
\subsection{Zero-shot Evaluation  (Q2)}
\paragraph{ProteinGym May Not Be Suitable for Evaluating PFMs.} Table~\ref{tab:zero_shot_proteingym} summarizes the zero-shot performance of core models on the ProteinGym benchmark, following the evaluation protocol outlined in ProteinGym \cite{notin2023proteingym}. Models such as ProtST, ProtoT5, and ProstT5 could not be evaluated under this protocol and were therefore excluded. For ESM3, we evaluated both sequence-only and sequence-structure inputs, finding that the sequence-only version performed better. Interestingly, ProteinGym performance does not correlate with supervised tuning results, challenging the assumption that zero-shot performance is a reliable indicator of supervised performance. For future PLM development, we recommend prioritizing the 11 representative supervised tasks over zero-shot ProteinGym. 

\vspace{-4mm}
\begin{table}[h]
    \centering
    \caption{Zero-shot proteingym performance of core models.}
    \label{tab:zero_shot_proteingym}
    \resizebox{0.7 \columnwidth}{!}{\begin{tabular}{lllllll}
    \toprule
             & \# Params & Architecture    & Input      & Loss                         & ProteinGym & Rank \\ \midrule
    SaProt \cite{susaprot}  & 650M      & Encoder         & Seq-Struct & MLM                      & 0.45094    & 1    \\
    VenusPLM \cite{tan2025venusfactory} & 300M      & Encoder         & Seq        & MLM                      & 0.43952    & 2    \\
    ESM-2 \cite{lin2023evolutionary}   & 650M      & Encoder         & Seq        & MLM                      & 0.43904    & 3    \\
    ESM-C    & 600M      & Encoder         & Seq        & MLM                      & 0.43422    & 4    \\
    DPLM  \cite{wang2024diffusion}    & 650M      & Encoder         & Seq        & MLM                      & 0.42922    & 5    \\
    ESM3 \cite{hayes2025simulating}    & 1.4B      & Encoder         & Seq & MLM               & 0.41401     & 6    \\
    PGLM  \cite{chen2024xtrimopglm}    & 1B        & Encoder-Decoder & Seq        & MLM                      & 0.39750     & 7   \\
    ProTrek \cite{su2024protrek} & 650M      & Encoder         & Seq        & MLM+ Contrast & 0.35919    & 8    \\
    ProtGPT2 \cite{ferruz2022protgpt2} & 738M      & Decoder         & Seq        & NTP                      & 0.18962    & 9   \\  \bottomrule
    \end{tabular}}
\end{table}
\vspace{-3mm}

\begin{figure}[h]
    \vspace{-4mm}
    \centering
    \includegraphics[width=3.5in]{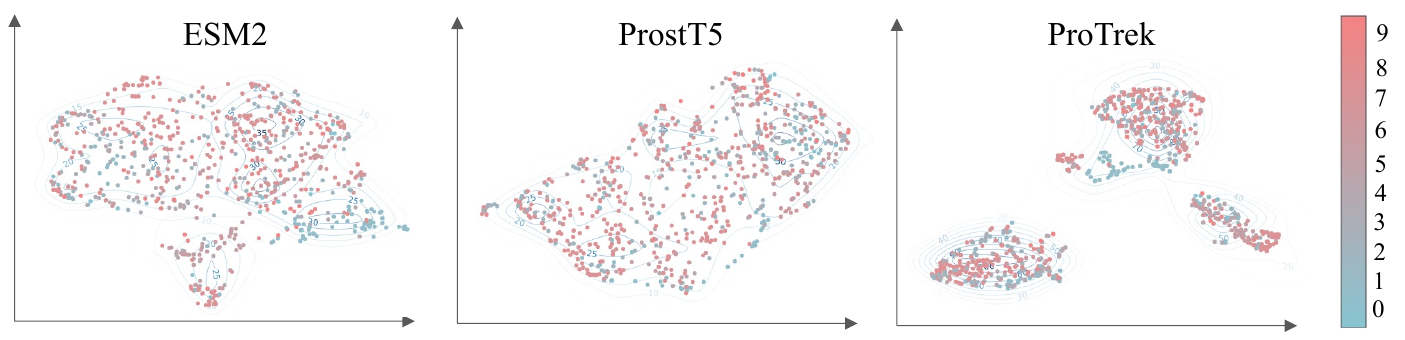}
   
    \caption{UMAP visualization of ESM2, ProstT5, and ProTrek on Deeploc2\_Multi.} 
    \label{fig:umap}
    \vspace{-3mm}
\end{figure}

\paragraph{UMAP Visualization.} Figure~\ref{fig:umap} shows UMAP embeddings of ESM2, ProstT5, and ProTrek on Deeploc2\_Multi, colored by class labels. ESM2 and ProstT5 exhibit overlapping clusters, while ProTrek, leveraging contrastive alignment, shows distinct boundaries. This highlights the importance of semantic alignment in pretraining for capturing functional relationships.
\vspace{-4mm}

\begin{figure}[h]
    \vspace{-3mm}
    \centering
    \includegraphics[width=4in]{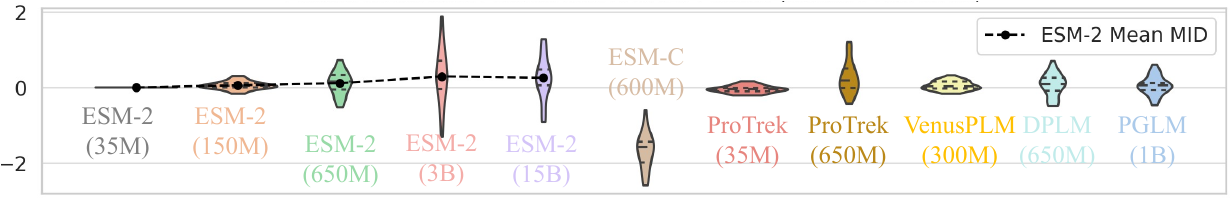}
    \caption{The MID distribution of sequence-only models relative to ESM2-35M.} 
    \label{fig:MID}
\end{figure}

\paragraph{MSA Mutual Information.} We compute the Mutual Information Difference (MID) for sequence-only models relative to ESM2-35M across 100 MSA clusters (see Appendix~\ref{appendix:MI} for MID definition). MSA centers are randomly sampled from UniRef30 \cite{suzek2015uniref}, with mmseq2 \cite{steinegger2017mmseqs2} used for top-10 MSA searches. Figure~\ref{fig:MID} shows that ProTrek and larger ESM models achieve higher MID, consistent with their downstream performance, suggesting that PLMs effectively clustering local MSA.

\vspace{-4mm}

\clearpage
\subsection{Optimal Efficient Fine-tuning and Scaling (Q3 \& Q4)}
Table~\ref{tab:core_model_performance_peft} presents the performance of the top-2 models alongside the ESM2 baseline on 11 representative tasks using various efficient fine-tuning methods, including Adapter, Linear Probing, LoRA, AdaLoRA, DoRA, and IA3. For each fine-tuning method, we calculate the winning rate (\#WESM) against ESM2. Additionally, across different fine-tuning methods, we compute the winning rate (\#WAdap) against the adapter tuning method for each model. We observe that:
\begin{itemize}[leftmargin=1.5em]
    \item \textbf{Adapter Tuning is Sufficiently Effective.} The adapter tuning method consistently outperforms other fine-tuning methods across all models, except for DoRA.
    \item \textbf{ProTrek Consistently Outperforms ESM2.} ProTrek achieves the best performance across all fine-tuning methods, with a winning rate of 75\% to 88\% against ESM2. 
\end{itemize}
\vspace{-5mm}
\begin{table}[h]
    \centering
    \caption{Results on 11 representative tasks using various efficient fine-tuning methods. PEFT methods that outperform the Adapter are marked in red; the others are marked in blue.}
    \label{tab:core_model_performance_peft}
    \resizebox{1.0 \columnwidth}{!}{\begin{tabular}{l
        >{\color{gray}}l
        >{\color{gray}}l
        >{\color{gray}}l
        lllllllllllll}
        \toprule
            & PDBBind & BindingDB & Stability & Anti. Res. & Mat. Prod. & EC      & M. I. Bin & Sec. Str. & DL2 Multi & Clo. CLF & DeepSol & \#WESM & \#WAdap \\ \midrule
    \multicolumn{12}{l}{\textbf{Adapter}}                                                                                                  \\ \shadeRow{50}{good}
    ESM-2   & 0.14677 & 0.13692   & 0.32112   & 0.63422    & 0.81189    & 0.73578 & 0.71170   & 0.76375   & 0.76191   & 0.80586  & 0.84494 &  -- &  --\\ \shadeRow{50}{good}
    ProstT5 & 0.18344 & 0.16642   & 0.13032   & \underline{0.69027}    & 0.81622    & \textbf{0.76829} & 0.72145   & \textbf{0.81397}   & 0.73190   & 0.79853  & 0.81937 &  63\% &  --\\ \shadeRow{0}{good}
    ProTrek & 0.17322 & 0.19230   & 0.04924  & 0.59292    & 0.81477    & \underline{0.76408} & \textbf{0.80362}   & 0.77363   & \textbf{0.83944}   & 0.82612  & 0.83427 &  75\% &  --\\ \midrule
    \multicolumn{12}{l}{\textbf{Linear Probing}}                                                                                           \\ \shadeRow{50}{bad}
    ESM-2   & 0.21766 & 0.16427   & 0.04649  & 0.64307    & 0.81586    & 0.61163 & 0.71309   & 0.71846   & 0.74472   & 0.78807  & 0.79465 &  -- & 38\%\\ \shadeRow{50}{bad}
    ProstT5 & \textbf{0.24874} & 0.16144   & 0.05228  & 0.67552    & 0.80541    & 0.65167 & 0.66992   & 0.79928   & 0.70530    & 0.78722  & 0.77794 &  38\% &  0\%\\ \shadeRow{50}{bad}
    ProTrek & 0.22595 & \textbf{0.25353}   & 0.02332  & 0.63717    & 0.81766    & 0.64201 & 0.69777   & 0.73840    & 0.77847   & 0.80099  & 0.80312 &  75\% &  25\%\\ \midrule
    \multicolumn{12}{l}{\textbf{LoRA}}                                                                                                     \\ \shadeRow{50}{bad}
    ESM-2   & 0.18463 & \underline{0.24559}   & \underline{0.32304}   & 0.61652    & 0.80865    & 0.67146 & 0.69499   & 0.74305   & 0.76851   & 0.83616  & 0.86160  &  --  & 38\%\\ \shadeRow{50}{bad}
    ProstT5 & 0.19072 & 0.21411   & 0.28204   & 0.66077    & 0.81658    & 0.72779 & 0.64485   & 0.80878   & 0.77875   & 0.82997  & 0.84834 &  63\% &  0\%\\ \shadeRow{50}{bad}
    ProTrek & \underline{0.24707} & 0.19302   & 0.2776    & 0.67257    & \textbf{0.84324}    & 0.71139 & \underline{0.74373}   & 0.76687   & 0.79566   & 0.83441  & 0.86326 &  88\% &  50\%\\ \midrule
    \multicolumn{12}{l}{\textbf{AdaLoRA}}                                                                                                  \\ 
    ESM-2   & 0.20398 & 0.23794   & 0.26715   & 0.60767    & 0.80829    & 0.68715 & 0.71448   & 0.7436    & 0.77209   & 0.84171  & 0.85077 &  --  & 50\%\\
    ProstT5 & 0.21487 & 0.07897   & 0.17776   & 0.68142    & 0.82883    & 0.71974 & 0.66156   & 0.80755   & 0.75642   & 0.82935  & 0.85272 &  63\% &  50\%\\
    ProTrek & 0.24625 & 0.22491   & 0.15328   & 0.64307    & \underline{0.83640}    & 0.7384  & 0.68524   & 0.76651   & {0.80497}   & 0.83713  & 0.86152 &  75\% &  50\%\\ \midrule
    \multicolumn{12}{l}{\textbf{DoRA}}                                                                                                     \\ \shadeRow{50}{good}
    ESM-2   & 0.18497 & 0.20087   & \textbf{0.33022}   & 0.63717    & 0.82739    & 0.68786 & 0.72006   & 0.74357   & 0.77774   & \textbf{0.84471}  & \underline{0.86346} &  --  & 75\%\\ \shadeRow{50}{good}
    ProstT5 & 0.23039 & 0.10505   & 0.26731   & \textbf{0.69912}    & 0.80000    & 0.70583 & 0.66574   & 0.80813   & 0.77520    & 0.83052  & 0.85343 &  38\% &  50\%\\ \shadeRow{0}{good}
    ProTrek & 0.23648 & 0.07242  & 0.25293   & 0.60177    & 0.83387    & 0.71772 & 0.72006   & 0.76710    & \underline{0.80063}          & \underline{0.83988}  & \textbf{0.86625} & 63\% & 75\% \\ \midrule
    \multicolumn{12}{l}{\textbf{IA3}}                                                                                                      \\ \shadeRow{50}{bad}
    ESM-2   & 0.18948 & 0.19144   & 0.09641  & 0.60177    & 0.79928    & 0.68549 & 0.63231   & 0.74286   & 0.76447   & 0.82562  & 0.83062 &  --  & 25\%\\ \shadeRow{50}{bad}
    ProstT5 & 0.24188 & 0.12700   & 0.04821  & 0.66962    & 0.82342    & 0.71467 & 0.71309   & \underline{0.81016}   & 0.74326   & 0.78942  & 0.80635 &  63\% &  25\%\\ \shadeRow{50}{bad}
    ProTrek & 0.23836 & 0.10734   & 0.06299  & 0.59292    & 0.79676    &  0.70588       & 0.71031   & 0.76366   & 0.78881   & 0.82911  & 0.83146 & 75\% & 25\%  \\ \bottomrule
    \end{tabular}}
\end{table}

\paragraph{Are Scaling PLMs Truly Worth It?} In Table~\ref{tab:core_model_performance_scale}, we further examine whether increasing model size improves performance on the 11 representative tasks, focusing on the ESM2 series models. We calculate the winning rate (W150M) of each model against ESM2-150M and conclude the following:
\begin{itemize}[leftmargin=1.5em]
    \item \textbf{Scaling Up Works but Comes at a Cost.} The scaling law is effective only when models are scaled up to 15B parameters; otherwise, none of the models outperform ESM2-150M. However, this increase in model size incurs significant costs in both pretraining and inference. Considering the marginal performance gains, the cost of scaling up may not be justified.
    \item \textbf{Pretraining Strategies Matter More.} Instead of scaling up to 15B, a more effective and efficient approach is to optimize the pretraining strategy. For instance, ProTrek-650M outperforms ESM2-15B on 6 out of 8 tasks and achieves a winning rate of 75\% against ESM2-150M.
\end{itemize}
\begin{table}[h]
    \small
    \caption{Performance of ESM2 under the scaling law. Gray tasks are excluded from the winning rate analysis.  Models that outperform the ESM2-150M are marked in red; the others are marked in blue. }
    \label{tab:core_model_performance_scale}
    \resizebox{1.0 \columnwidth}{!}{\begin{tabular}{l
        >{\color{gray}}l
        >{\color{gray}}l
        >{\color{gray}}l
        llllllllllll} \toprule
                     & PDBBind          & Bind. DB    & Stability      & Anti.Res.        & Mat.Pro.         & EC               & M. I. Bin.    & Sec. Str.        & DL2 M.    & Clo. CLF        & DeepSol     & \#W150M  \\ \midrule \shadeRow{50}{bad}
    ESM2-35M    & 0.09985 & \underline{0.14232}   & \underline{0.32337}        & \underline{0.67552}   & 0.78595   & 0.71675 & 0.71866    & 0.69609   & 0.73219   & 0.79441  & 0.82486 & 13\%\\
    ESM2-150M   & 0.09371 & 0.13142   & \textbf{0.33728}         & 0.65192   & \textbf{0.81946}   & 0.73192 & \underline{0.76462}    & 0.73430   & 0.74744   & \underline{0.81531}  & 0.82825 & -- \\
    ESM2-650M            & \underline{0.14677}          & 0.13692   & 0.32112       & 0.63422          & 0.81189          & 0.73578          & 0.71170          & 0.76375          & { 0.76191}    & 0.80586          & \underline{0.84494} & 50\%\\
    ESM2-3B     & 0.10479 & 0.12724   & 0.31647         & 0.64012   & 0.80036   & 0.73878 & 0.73955    & 0.77111   & 0.77328   & 0.81031  & 0.83007  & 50\%\\ \shadeRow{70}{good}
    ESM2-15B    & 0.08427 & 0.12559   & 0.03018         & \textbf{0.68142}   & 0.81045   & 0.73259 & 0.73259    & \underline{0.77250}    & 0.76714   & 0.80210  & \textbf{0.85155} & 63\% \\ 
    \midrule \shadeRow{50}{good}
    ProTrek-650M         & \textbf{0.17322}          & \textbf{0.19230}   & 0.04924       & 0.59292          & \underline{0.81477}          & \textbf{0.76408}    & \textbf{0.80362} & \textbf{0.77363}          & \textbf{0.83944} & \textbf{0.82612}          & 0.83427 & 75\%  \\ 
    \bottomrule       
    \end{tabular}}
\end{table}

\vspace{-3mm}
\section{Conclusion}
\vspace{-3mm}
This work presents a comprehensive benchmark for evaluating protein foundation models (PFMs) across a diverse range of tasks, accompanied by a streamlined evaluation protocol. Starting with 38 tasks and 17 models, we identify 12 core models and 11 representative tasks to enable efficient and meaningful evaluation. Through extensive experiments, we reveal that current PFM research exhibits a high degree of homogeneity and provide in-depth analysis to guide future research directions.


\bibliographystyle{plain}
\bibliography{citation}

\begin{thebibliography}{10}

\bibitem{almagro2017deeploc}
Jos{\'e}~Juan Almagro~Armenteros, Casper~Kaae S{\o}nderby, S{\o}ren~Kaae S{\o}nderby, Henrik Nielsen, and Ole Winther.
\newblock Deeploc: prediction of protein subcellular localization using deep learning.
\newblock {\em Bioinformatics}, 33(21):3387--3395, 2017.

\bibitem{ashburner2000gene}
Michael Ashburner, Catherine~A Ball, Judith~A Blake, David Botstein, Heather Butler, J~Michael Cherry, Allan~P Davis, Kara Dolinski, Selina~S Dwight, Janan~T Eppig, et~al.
\newblock Gene ontology: tool for the unification of biology.
\newblock {\em Nature genetics}, 25(1):25--29, 2000.

\bibitem{bairoch2000enzyme}
Amos Bairoch.
\newblock The enzyme database in 2000.
\newblock {\em Nucleic acids research}, 28(1):304--305, 2000.

\bibitem{bjerregaard2025foundation}
Andreas Bjerregaard, Peter~M{\o}rch Groth, S{\o}ren Hauberg, Anders Krogh, and Wouter Boomsma.
\newblock Foundation models of protein sequences: A brief overview.
\newblock {\em Current Opinion in Structural Biology}, 91:103004, 2025.

\bibitem{chen2024xtrimopglm}
Bo~Chen, Xingyi Cheng, Pan Li, Yangli-ao Geng, Jing Gong, Shen Li, Zhilei Bei, Xu~Tan, Boyan Wang, Xin Zeng, et~al.
\newblock xtrimopglm: unified 100b-scale pre-trained transformer for deciphering the language of protein.
\newblock {\em arXiv preprint arXiv:2401.06199}, 2024.

\bibitem{chen2021structure}
Jianwen Chen, Shuangjia Zheng, Huiying Zhao, and Yuedong Yang.
\newblock Structure-aware protein solubility prediction from sequence through graph convolutional network and predicted contact map.
\newblock {\em Journal of cheminformatics}, 13:1--10, 2021.

\bibitem{dallago2flip}
Christian Dallago, Jody Mou, Kadina~E Johnston, Bruce Wittmann, Nick Bhattacharya, Samuel Goldman, Ali Madani, and Kevin~K Yang.
\newblock Flip: Benchmark tasks in fitness landscape inference for proteins.
\newblock In {\em Thirty-fifth Conference on Neural Information Processing Systems Datasets and Benchmarks Track (Round 2)}.

\bibitem{dettmers2023qlora}
Tim Dettmers, Artidoro Pagnoni, Ari Holtzman, and Luke Zettlemoyer.
\newblock Qlora: Efficient finetuning of quantized llms.
\newblock {\em Advances in neural information processing systems}, 36:10088--10115, 2023.

\bibitem{du2025loca}
Zhekai Du, Yinjie Min, Jingjing Li, Ke~Lu, Changliang Zou, Liuhua Peng, Tingjin Chu, and Mingming Gong.
\newblock Loca: Location-aware cosine adaptation for parameter-efficient fine-tuning.
\newblock {\em arXiv preprint arXiv:2502.06820}, 2025.

\bibitem{elnaggar2023ankh}
Ahmed Elnaggar, Hazem Essam, Wafaa Salah-Eldin, Walid Moustafa, Mohamed Elkerdawy, Charlotte Rochereau, and Burkhard Rost.
\newblock Ankh: Optimized protein language model unlocks general-purpose modelling.
\newblock {\em arXiv preprint arXiv:2301.06568}, 2023.

\bibitem{elnaggar2021prottrans}
Ahmed Elnaggar, Michael Heinzinger, Christian Dallago, Ghalia Rehawi, Yu~Wang, Llion Jones, Tom Gibbs, Tamas Feher, Christoph Angerer, Martin Steinegger, et~al.
\newblock Prottrans: towards cracking the language of life’s code through self-supervised learning.
\newblock {\em IEEE Transactions on Pattern Analysis and Machine Intelligence}, 44:7112--7127, 2021.

\bibitem{ferruz2022protgpt2}
Noelia Ferruz, Steffen Schmidt, and Birte H{\"o}cker.
\newblock Protgpt2 is a deep unsupervised language model for protein design.
\newblock {\em Nature communications}, 13(1):4348, 2022.

\bibitem{gado2023deep}
Japheth~E Gado, Matthew Knotts, Ada~Y Shaw, Debora Marks, Nicholas~P Gauthier, Chris Sander, and Gregg~T Beckham.
\newblock Deep learning prediction of enzyme optimum ph.
\newblock {\em bioRxiv}, pages 2023--06, 2023.

\bibitem{gao2023proteininvbench}
Zhangyang Gao, Cheng Tan, Yijie Zhang, Xingran Chen, Lirong Wu, and Stan~Z Li.
\newblock Proteininvbench: Benchmarking protein inverse folding on diverse tasks, models, and metrics.
\newblock {\em Advances in Neural Information Processing Systems}, 36:68207--68220, 2023.

\bibitem{gray2018quantitative}
Vanessa~E Gray, Ronald~J Hause, Jens Luebeck, Jay Shendure, and Douglas~M Fowler.
\newblock Quantitative missense variant effect prediction using large-scale mutagenesis data.
\newblock {\em Cell systems}, 6(1):116--124, 2018.

\bibitem{guo2025foundation}
Fei Guo, Renchu Guan, Yaohang Li, Qi~Liu, Xiaowo Wang, Can Yang, and Jianxin Wang.
\newblock Foundation models in bioinformatics.
\newblock {\em National Science Review}, page nwaf028, 2025.

\bibitem{guo2008using}
Yanzhi Guo, Lezheng Yu, Zhining Wen, and Menglong Li.
\newblock Using support vector machine combined with auto covariance to predict protein--protein interactions from protein sequences.
\newblock {\em Nucleic acids research}, 36(9):3025--3030, 2008.

\bibitem{hayes2025simulating}
Thomas Hayes, Roshan Rao, Halil Akin, Nicholas~J Sofroniew, Deniz Oktay, Zeming Lin, Robert Verkuil, Vincent~Q Tran, Jonathan Deaton, Marius Wiggert, et~al.
\newblock Simulating 500 million years of evolution with a language model.
\newblock {\em Science}, page eads0018, 2025.

\bibitem{hetowards}
Junxian He, Chunting Zhou, Xuezhe Ma, Taylor Berg-Kirkpatrick, and Graham Neubig.
\newblock Towards a unified view of parameter-efficient transfer learning.
\newblock In {\em International Conference on Learning Representations}.

\bibitem{he2016deep}
Kaiming He, Xiangyu Zhang, Shaoqing Ren, and Jian Sun.
\newblock Deep residual learning for image recognition.
\newblock In {\em Proceedings of the IEEE conference on computer vision and pattern recognition}, pages 770--778, 2016.

\bibitem{heinzinger2024bilingual}
Michael Heinzinger, Konstantin Weissenow, Joaquin~Gomez Sanchez, Adrian Henkel, Milot Mirdita, Martin Steinegger, and Burkhard Rost.
\newblock Bilingual language model for protein sequence and structure.
\newblock {\em NAR Genomics and Bioinformatics}, 6(4):lqae150, 2024.

\bibitem{hochreiter1997long}
Sepp Hochreiter and J{\"u}rgen Schmidhuber.
\newblock Long short-term memory.
\newblock {\em Neural computation}, 9(8):1735--1780, 1997.

\bibitem{houlsby2019parameter}
Neil Houlsby, Andrei Giurgiu, Stanislaw Jastrzebski, Bruna Morrone, Quentin De~Laroussilhe, Andrea Gesmundo, Mona Attariyan, and Sylvain Gelly.
\newblock Parameter-efficient transfer learning for nlp.
\newblock In {\em International conference on machine learning}, pages 2790--2799. PMLR, 2019.

\bibitem{hu2022lora}
Edward~J Hu, Yelong Shen, Phillip Wallis, Zeyuan Allen-Zhu, Yuanzhi Li, Shean Wang, Lu~Wang, Weizhu Chen, et~al.
\newblock Lora: Low-rank adaptation of large language models.
\newblock {\em ICLR}, 1(2):3, 2022.

\bibitem{hu2022exploring}
Mingyang Hu, Fajie Yuan, Kevin Yang, Fusong Ju, Jin Su, Hui Wang, Fei Yang, and Qiuyang Ding.
\newblock Exploring evolution-aware \&-free protein language models as protein function predictors.
\newblock {\em Advances in Neural Information Processing Systems}, 35:38873--38884, 2022.

\bibitem{jarzab2020meltome}
Anna Jarzab, Nils Kurzawa, Thomas Hopf, Matthias Moerch, Jana Zecha, Niels Leijten, Yangyang Bian, Eva Musiol, Melanie Maschberger, Gabriele Stoehr, et~al.
\newblock Meltome atlas—thermal proteome stability across the tree of life.
\newblock {\em Nature methods}, 17(5):495--503, 2020.

\bibitem{khurana2018deepsol}
Sameer Khurana, Reda Rawi, Khalid Kunji, Gwo-Yu Chuang, Halima Bensmail, and Raghvendra Mall.
\newblock Deepsol: a deep learning framework for sequence-based protein solubility prediction.
\newblock {\em Bioinformatics}, 34(15):2605--2613, 2018.

\bibitem{klausen2019netsurfp}
Michael~Schantz Klausen, Martin~Closter Jespersen, Henrik Nielsen, Kamilla~Kjaergaard Jensen, Vanessa~Isabell Jurtz, Casper~Kaae Soenderby, Morten Otto~Alexander Sommer, Ole Winther, Morten Nielsen, Bent Petersen, et~al.
\newblock Netsurfp-2.0: Improved prediction of protein structural features by integrated deep learning.
\newblock {\em Proteins: Structure, Function, and Bioinformatics}, 87(6):520--527, 2019.

\bibitem{koyama2023attention}
Kyohei Koyama, Kosuke Hashimoto, Chioko Nagao, and Kenji Mizuguchi.
\newblock Attention network for predicting t-cell receptor--peptide binding can associate attention with interpretable protein structural properties.
\newblock {\em Frontiers in Bioinformatics}, 3:1274599, 2023.

\bibitem{lester2021power}
Brian Lester, Rami Al-Rfou, and Noah Constant.
\newblock The power of scale for parameter-efficient prompt tuning.
\newblock In {\em Proceedings of the 2021 Conference on Empirical Methods in Natural Language Processing}, pages 3045--3059, 2021.

\bibitem{li2022deep}
Feiran Li, Le~Yuan, Hongzhong Lu, Gang Li, Yu~Chen, Martin~KM Engqvist, Eduard~J Kerkhoven, and Jens Nielsen.
\newblock Deep learning-based k cat prediction enables improved enzyme-constrained model reconstruction.
\newblock {\em Nature Catalysis}, 5(8):662--672, 2022.

\bibitem{li2022learning}
Gang Li, Filip Buric, Jan Zrimec, Sandra Viknander, Jens Nielsen, Aleksej Zelezniak, and Martin~KM Engqvist.
\newblock Learning deep representations of enzyme thermal adaptation.
\newblock {\em Protein Science}, 31(12):e4480, 2022.

\bibitem{li2024progress}
Qing Li, Zhihang Hu, Yixuan Wang, Lei Li, Yimin Fan, Irwin King, Gengjie Jia, Sheng Wang, Le~Song, and Yu~Li.
\newblock Progress and opportunities of foundation models in bioinformatics.
\newblock {\em Briefings in Bioinformatics}, 25(6):bbae548, 2024.

\bibitem{li2021prefix}
Xiang~Lisa Li and Percy Liang.
\newblock Prefix-tuning: Optimizing continuous prompts for generation.
\newblock In {\em Proceedings of the 59th Annual Meeting of the Association for Computational Linguistics and the 11th International Joint Conference on Natural Language Processing (Volume 1: Long Papers)}, pages 4582--4597, 2021.

\bibitem{lin2023evolutionary}
Zeming Lin, Halil Akin, Roshan Rao, Brian Hie, Zhongkai Zhu, Wenting Lu, Nikita Smetanin, Robert Verkuil, Ori Kabeli, Yaniv Shmueli, et~al.
\newblock Evolutionary-scale prediction of atomic-level protein structure with a language model.
\newblock {\em Science}, 379(6637):1123--1130, 2023.

\bibitem{liu2022few}
Haokun Liu, Derek Tam, Mohammed Muqeeth, Jay Mohta, Tenghao Huang, Mohit Bansal, and Colin~A Raffel.
\newblock Few-shot parameter-efficient fine-tuning is better and cheaper than in-context learning.
\newblock {\em Advances in Neural Information Processing Systems}, 35:1950--1965, 2022.

\bibitem{liu2007bindingdb}
Tiqing Liu, Yuhmei Lin, Xin Wen, Robert~N Jorissen, and Michael~K Gilson.
\newblock Bindingdb: a web-accessible database of experimentally determined protein--ligand binding affinities.
\newblock {\em Nucleic acids research}, 35(suppl\_1):D198--D201, 2007.

\bibitem{liu2017forging}
Zhihai Liu, Minyi Su, Li~Han, Jie Liu, Qifan Yang, Yan Li, and Renxiao Wang.
\newblock Forging the basis for developing protein--ligand interaction scoring functions.
\newblock {\em Accounts of chemical research}, 50(2):302--309, 2017.

\bibitem{lo2000scop}
Loredana Lo~Conte, Bart Ailey, Tim~JP Hubbard, Steven~E Brenner, Alexey~G Murzin, and Cyrus Chothia.
\newblock Scop: a structural classification of proteins database.
\newblock {\em Nucleic acids research}, 28(1):257--259, 2000.

\bibitem{lv2025prollama}
Liuzhenghao Lv, Zongying Lin, Hao Li, Yuyang Liu, Jiaxi Cui, Calvin Yu-Chian Chen, Li~Yuan, and Yonghong Tian.
\newblock Prollama: A protein large language model for multi-task protein language processing.
\newblock {\em IEEE Transactions on Artificial Intelligence}, 2025.

\bibitem{madani2023large}
Ali Madani, Ben Krause, Eric~R Greene, Subu Subramanian, Benjamin~P Mohr, James~M Holton, Jose~Luis Olmos~Jr, Caiming Xiong, Zachary~Z Sun, Richard Socher, et~al.
\newblock Large language models generate functional protein sequences across diverse families.
\newblock {\em Nature biotechnology}, 41(8):1099--1106, 2023.

\bibitem{peft}
Sourab Mangrulkar, Sylvain Gugger, Lysandre Debut, Younes Belkada, Sayak Paul, and Benjamin Bossan.
\newblock Peft: State-of-the-art parameter-efficient fine-tuning methods.
\newblock \url{https://github.com/huggingface/peft}, 2022.

\bibitem{mao2024dora}
Yulong Mao, Kaiyu Huang, Changhao Guan, Ganglin Bao, Fengran Mo, and Jinan Xu.
\newblock Dora: Enhancing parameter-efficient fine-tuning with dynamic rank distribution.
\newblock In {\em Proceedings of the 62nd Annual Meeting of the Association for Computational Linguistics (Volume 1: Long Papers)}, pages 11662--11675, 2024.

\bibitem{mcallester2020formal}
David McAllester and Karl Stratos.
\newblock Formal limitations on the measurement of mutual information.
\newblock In {\em International Conference on Artificial Intelligence and Statistics}, pages 875--884. PMLR, 2020.

\bibitem{moal2012skempi}
Iain~H Moal and Juan Fern{\'a}ndez-Recio.
\newblock Skempi: a structural kinetic and energetic database of mutant protein interactions and its use in empirical models.
\newblock {\em Bioinformatics}, 28(20):2600--2607, 2012.

\bibitem{nijkamp2023progen2}
Erik Nijkamp, Jeffrey~A Ruffolo, Eli~N Weinstein, Nikhil Naik, and Ali Madani.
\newblock Progen2: exploring the boundaries of protein language models.
\newblock {\em Cell systems}, 14(11):968--978, 2023.

\bibitem{notin2023proteingym}
Pascal Notin, Aaron Kollasch, Daniel Ritter, Lood Van~Niekerk, Steffanie Paul, Han Spinner, Nathan Rollins, Ada Shaw, Rose Orenbuch, Ruben Weitzman, et~al.
\newblock Proteingym: Large-scale benchmarks for protein fitness prediction and design.
\newblock {\em Advances in Neural Information Processing Systems}, 36:64331--64379, 2023.

\bibitem{pan2010large}
Xiao-Yong Pan, Ya-Nan Zhang, and Hong-Bin Shen.
\newblock Large-scale prediction of human protein- protein interactions from amino acid sequence based on latent topic features.
\newblock {\em Journal of proteome research}, 9(10):4992--5001, 2010.

\bibitem{pfeiffer2020adapterfusion}
Jonas Pfeiffer, Aishwarya Kamath, Andreas R{\"u}ckl{\'e}, Kyunghyun Cho, and Iryna Gurevych.
\newblock Adapterfusion: Non-destructive task composition for transfer learning.
\newblock {\em arXiv preprint arXiv:2005.00247}, 2020.

\bibitem{poole2019variational}
Ben Poole, Sherjil Ozair, Aaron Van Den~Oord, Alex Alemi, and George Tucker.
\newblock On variational bounds of mutual information.
\newblock In {\em International conference on machine learning}, pages 5171--5180. PMLR, 2019.

\bibitem{queen2024procyon}
Owen Queen, Yepeng Huang, Robert Calef, Valentina Giunchiglia, Tianlong Chen, George Dasoulas, LeAnn Tai, Yasha Ektefaie, Ayush Noori, Joseph Brown, et~al.
\newblock Procyon: A multimodal foundation model for protein phenotypes.
\newblock {\em BioRxiv}, pages 2024--12, 2024.

\bibitem{rao2019evaluating}
Roshan Rao, Nicholas Bhattacharya, Neil Thomas, Yan Duan, Peter Chen, John Canny, Pieter Abbeel, and Yun Song.
\newblock Evaluating protein transfer learning with tape.
\newblock {\em Advances in neural information processing systems}, 32, 2019.

\bibitem{rives2021biological}
Alexander Rives, Joshua Meier, Tom Sercu, Siddharth Goyal, Zeming Lin, Jason Liu, Demi Guo, Myle Ott, C~Lawrence Zitnick, Jerry Ma, et~al.
\newblock Biological structure and function emerge from scaling unsupervised learning to 250 million protein sequences.
\newblock {\em Proceedings of the National Academy of Sciences}, 118(15):e2016239118, 2021.

\bibitem{steinegger2017mmseqs2}
Martin Steinegger and Johannes S{\"o}ding.
\newblock Mmseqs2 enables sensitive protein sequence searching for the analysis of massive data sets.
\newblock {\em Nature biotechnology}, 35(11):1026--1028, 2017.

\bibitem{susaprot}
Jin Su, Chenchen Han, Yuyang Zhou, Junjie Shan, Xibin Zhou, and Fajie Yuan.
\newblock Saprot: Protein language modeling with structure-aware vocabulary.
\newblock In {\em The Twelfth International Conference on Learning Representations}.

\bibitem{su2024protrek}
Jin Su, Xibin Zhou, Xuting Zhang, and Fajie Yuan.
\newblock Protrek: Navigating the protein universe through tri-modal contrastive learning.
\newblock {\em bioRxiv}, pages 2024--05, 2024.

\bibitem{suzek2015uniref}
Baris~E Suzek, Yuqi Wang, Hongzhan Huang, Peter~B McGarvey, Cathy~H Wu, and UniProt Consortium.
\newblock Uniref clusters: a comprehensive and scalable alternative for improving sequence similarity searches.
\newblock {\em Bioinformatics}, 31(6):926--932, 2015.

\bibitem{tan2024peta}
Yang Tan, Mingchen Li, Ziyi Zhou, Pan Tan, Huiqun Yu, Guisheng Fan, and Liang Hong.
\newblock Peta: evaluating the impact of protein transfer learning with sub-word tokenization on downstream applications.
\newblock {\em Journal of Cheminformatics}, 16(1):92, 2024.

\bibitem{tan2025venusfactory}
Yang Tan, Chen Liu, Jingyuan Gao, Banghao Wu, Mingchen Li, Ruilin Wang, Lingrong Zhang, Huiqun Yu, Guisheng Fan, Liang Hong, et~al.
\newblock Venusfactory: A unified platform for protein engineering data retrieval and language model fine-tuning.
\newblock {\em arXiv preprint arXiv:2503.15438}, 2025.

\bibitem{tan2024protsolm}
Yang Tan, Jia Zheng, Liang Hong, and Bingxin Zhou.
\newblock Protsolm: Protein solubility prediction with multi-modal features.
\newblock In {\em 2024 IEEE International Conference on Bioinformatics and Biomedicine (BIBM)}, pages 223--232. IEEE, 2024.

\bibitem{thumuluri2022deeploc}
Vineet Thumuluri, Jos{\'e}~Juan Almagro~Armenteros, Alexander~Rosenberg Johansen, Henrik Nielsen, and Ole Winther.
\newblock Deeploc 2.0: multi-label subcellular localization prediction using protein language models.
\newblock {\em Nucleic acids research}, 50(W1):W228--W234, 2022.

\bibitem{tschannenmutual}
Michael Tschannen, Josip Djolonga, Paul~K Rubenstein, Sylvain Gelly, and Mario Lucic.
\newblock On mutual information maximization for representation learning.
\newblock In {\em International Conference on Learning Representations}.

\bibitem{varadi2022alphafold}
Mihaly Varadi, Stephen Anyango, Mandar Deshpande, Sreenath Nair, Cindy Natassia, Galabina Yordanova, David Yuan, Oana Stroe, Gemma Wood, Agata Laydon, et~al.
\newblock Alphafold protein structure database: massively expanding the structural coverage of protein-sequence space with high-accuracy models.
\newblock {\em Nucleic acids research}, 50(D1):D439--D444, 2022.

\bibitem{vaswani2017attention}
Ashish Vaswani, Noam Shazeer, Niki Parmar, Jakob Uszkoreit, Llion Jones, Aidan~N Gomez, {\L}ukasz Kaiser, and Illia Polosukhin.
\newblock Attention is all you need.
\newblock {\em NeurIPS}, 30, 2017.

\bibitem{wang2023prediction}
Chao Wang and Quan Zou.
\newblock Prediction of protein solubility based on sequence physicochemical patterns and distributed representation information with deepsolue.
\newblock {\em BMC biology}, 21(1):12, 2023.

\bibitem{wang2014predppcrys}
Huilin Wang, Mingjun Wang, Hao Tan, Yuan Li, Ziding Zhang, and Jiangning Song.
\newblock Predppcrys: accurate prediction of sequence cloning, protein production, purification and crystallization propensity from protein sequences using multi-step heterogeneous feature fusion and selection.
\newblock {\em PloS one}, 9(8):e105902, 2014.

\bibitem{wang2024comprehensive}
Wenkang Wang, Yunyan Shuai, Qiurong Yang, Fuhao Zhang, Min Zeng, and Min Li.
\newblock A comprehensive computational benchmark for evaluating deep learning-based protein function prediction approaches.
\newblock {\em Briefings in Bioinformatics}, 25(2):bbae050, 2024.

\bibitem{wang2024diffusion}
Xinyou Wang, Zaixiang Zheng, Fei Ye, Dongyu Xue, Shujian Huang, and Quanquan Gu.
\newblock Diffusion language models are versatile protein learners.
\newblock In {\em International Conference on Machine Learning}, pages 52309--52333. PMLR, 2024.

\bibitem{wu2024mixture}
Xun Wu, Shaohan Huang, and Furu Wei.
\newblock Mixture of lora experts.
\newblock {\em arXiv preprint arXiv:2404.13628}, 2024.

\bibitem{wu2023ccbhla}
Yejian Wu, Lujing Cao, Zhipeng Wu, Xinyi Wu, Xinqiao Wang, and Hongliang Duan.
\newblock Ccbhla: pan-specific peptide--hla class i binding prediction via convolutional and bilstm features.
\newblock {\em bioRxiv}, pages 2023--04, 2023.

\bibitem{xu2023protst}
Minghao Xu, Xinyu Yuan, Santiago Miret, and Jian Tang.
\newblock Protst: Multi-modality learning of protein sequences and biomedical texts.
\newblock In {\em International Conference on Machine Learning}, pages 38749--38767. PMLR, 2023.

\bibitem{xu2022peer}
Minghao Xu, Zuobai Zhang, Jiarui Lu, Zhaocheng Zhu, Yangtian Zhang, Ma~Chang, Runcheng Liu, and Jian Tang.
\newblock Peer: a comprehensive and multi-task benchmark for protein sequence understanding.
\newblock {\em Advances in Neural Information Processing Systems}, 35:35156--35173, 2022.

\bibitem{yang2024care}
Jason Yang, Ariane Mora, Shengchao Liu, Bruce Wittmann, Animashree Anandkumar, Frances Arnold, and Yisong Yue.
\newblock Care: a benchmark suite for the classification and retrieval of enzymes.
\newblock {\em Advances in Neural Information Processing Systems}, 37:3094--3121, 2024.

\bibitem{yang2020improved}
Jianyi Yang, Ivan Anishchenko, Hahnbeom Park, Zhenling Peng, Sergey Ovchinnikov, and David Baker.
\newblock Improved protein structure prediction using predicted interresidue orientations.
\newblock {\em Proceedings of the National Academy of Sciences}, 117(3):1496--1503, 2020.

\bibitem{ye2024proteinbench}
Fei Ye, Zaixiang Zheng, Dongyu Xue, Yuning Shen, Lihao Wang, Yiming Ma, Yan Wang, Xinyou Wang, Xiangxin Zhou, and Quanquan Gu.
\newblock Proteinbench: A holistic evaluation of protein foundation models.
\newblock {\em arXiv preprint arXiv:2409.06744}, 2024.

\bibitem{zaken2022bitfit}
Elad~Ben Zaken, Yoav Goldberg, and Shauli Ravfogel.
\newblock Bitfit: Simple parameter-efficient fine-tuning for transformer-based masked language-models.
\newblock In {\em Proceedings of the 60th Annual Meeting of the Association for Computational Linguistics}, pages 1--9, 2022.

\bibitem{zhangontoprotein}
Ningyu Zhang, Zhen Bi, Xiaozhuan Liang, Siyuan Cheng, Haosen Hong, Shumin Deng, Qiang Zhang, Jiazhang Lian, and Huajun Chen.
\newblock Ontoprotein: Protein pretraining with gene ontology embedding.
\newblock In {\em International Conference on Learning Representations}.

\bibitem{zhang2023adalora}
Qingru Zhang, Minshuo Chen, Alexander Bukharin, Nikos Karampatziakis, Pengcheng He, Yu~Cheng, Weizhu Chen, and Tuo Zhao.
\newblock Adalora: Adaptive budget allocation for parameter-efficient fine-tuning.
\newblock {\em arXiv preprint arXiv:2303.10512}, 2023.

\bibitem{zhangprotein}
Zuobai Zhang, Minghao Xu, Arian~Rokkum Jamasb, Vijil Chenthamarakshan, Aurelie Lozano, Payel Das, and Jian Tang.
\newblock Protein representation learning by geometric structure pretraining.
\newblock In {\em The Eleventh International Conference on Learning Representations}.

\bibitem{zhou2025protclip}
Hanjing Zhou, Mingze Yin, Wei Wu, Mingyang Li, Kun Fu, Jintai Chen, Jian Wu, and Zheng Wang.
\newblock Protclip: Function-informed protein multi-modal learning.
\newblock In {\em Proceedings of the AAAI Conference on Artificial Intelligence}, volume~39, pages 22937--22945, 2025.

\end{thebibliography}


\appendix
\clearpage
\section{Appendix}
\subsection{Supported Tasks}
\label{sec:appendix_tasks}
\subsubsection*{Task1: Annotation} 
\textbf{(Definition \& Metric)} Annotation tasks aim to predict functional characteristics of proteins. These tasks include predicting the subcellular localization of proteins (Cellular Component), their biochemical activities (Molecular Function), the broader biological processes they participate in (Biological Process) \cite{ashburner2000gene}, and their classification number according to the chemical reactions they catalyze (Enzyme Commission) \cite{bairoch2000enzyme}. F1 Score is the primary metric.

\textbf{(Impact)} Accurate annotation facilitates the identification of protein roles in cellular contexts, aiding in the discovery of novel drug targets and the elucidation of disease pathways.

\subsubsection*{Task2: Solubility} 
\textbf{(Definition \& Metric)} Solubility tasks evaluate a protein's ability to remain soluble under physiological conditions, which is a critical factor for successful protein expression and purification. PFMBench includes datasets such as DeepSol \cite{khurana2018deepsol}, DeepSoluE \cite{wang2023prediction}, ProtSolM \cite{tan2024protsolm}, and eSOL \cite{chen2021structure}. The primary metrics are AUROC for DeepSoluE, DeepSol, and ProtSolM, and Spearman correlation for eSOL.

\textbf{(Impact)} Predicting protein solubility is crucial for the successful expression and purification of recombinant proteins, which are essential in drug development and industrial applications. Insoluble proteins can lead to aggregation, reducing biological activity and complicating downstream processes. 

\subsubsection*{Task3: Localization}
\textbf{(Definition \& Metric)} Localization tasks focus on predicting the specific subcellular compartments where proteins are localized, which is crucial for understanding protein functions and interaction networks. These tasks include DeepLoc Multi \cite{almagro2017deeploc}, DeepLoc2 Multi \cite{thumuluri2022deeploc}, DeepLoc Binary \cite{almagro2017deeploc}, and Sorting Signal \cite{thumuluri2022deeploc}. The evaluation metrics are Accuracy for DeepLoc Multi, F1 Score for DeepLoc2 Multi and Sorting Signal, and AUROC for DeepLoc Binary.

\textbf{(Impact)} Accurate localization prediction aids in deciphering protein functions, interactions, and cellular pathways, contributing to our understanding of cellular organization and dynamics.

\subsubsection*{Task4: Mutation}
\textbf{(Definition \& Metric)} Mutation tasks evaluate the impact of amino acid substitutions on protein properties, which is pivotal in understanding disease mechanisms and guiding protein engineering. PFMBench includes datasets such as PETA\_CHS\_Sol, PETA\_LGK\_Sol, PETA\_TEM\_Sol \cite{tan2024peta}, FLIP\_AAV, FLIP\_GB1 \cite{dallago2flip}, TAPE\_Stability, TAPE\_Fluorescence \cite{rao2019evaluating}, and $\beta$-lactamase activity \cite{gray2018quantitative}. The primary metric is Spearman correlation for all datasets.

\textbf{(Impact)} Understanding the effects of mutations on protein function and stability is vital for elucidating disease mechanisms and guiding therapeutic interventions.

\subsubsection*{Task5: Interaction}
\textbf{(Definition \& Metric)} Protein-protein and protein-ligand interactions are fundamental to cellular processes and drug discovery. These tasks include datasets such as Human-PPI \cite{pan2010large}, Yeast-PPI \cite{guo2008using}, PPI affinity \cite{moal2012skempi}, PDBbind \cite{liu2017forging}, BindingDB \cite{liu2007bindingdb}, Metal Ion Binding \cite{hu2022exploring}, Peptide HLA MHC Affinity \cite{wu2023ccbhla}, and TCR PMHC Affinity \cite{koyama2023attention}. The evaluation metrics include AUROC for Human-PPI, Yeast-PPI, Peptide HLA MHC Affinity, and TCR PMHC Affinity; Spearman correlation for PPI affinity, PDBbind, and BindingDB; and Accuracy for Metal Ion Binding.

\textbf{(Impact)}  Accurate interaction prediction is crucial for understanding cellular signaling pathways, protein complexes, and drug-target interactions, facilitating drug discovery and development.

\subsubsection*{Task6: Structure}
\textbf{(Definition \& Metric)} Structure tasks focus on predicting the structural properties of proteins based on their sequences, which is essential for understanding their function and stability. These tasks include Contact prediction \cite{yang2020improved}, Fold classification \cite{lo2000scop}, and Secondary structure prediction \cite{klausen2019netsurfp}. The evaluation metrics are Top L/5 for Contact prediction, and Accuracy for both Fold classification and Secondary structure prediction.

\textbf{(Impact)}  Accurate structure prediction enables the understanding of protein mechanisms, the design of novel proteins, and the development of structure-based therapeutics.

\subsubsection*{Task7: Production}
\textbf{(Definition \& Metric)} Production tasks involve predicting properties that influence protein expression and manufacturing, which are critical for biotechnological applications. Datasets include Optimal pH \cite{gado2023deep}, DeepET\_Topt \cite{li2022learning}, Cloning CLF, Material Production \cite{wang2014predppcrys}, Enzyme Catalytic Efficiency \cite{li2022deep}, Antibiotic Resistance \cite{hu2022exploring}, and Thermostability \cite{jarzab2020meltome}. The evaluation metrics include Spearman correlation for Optimal pH, Enzyme Catalytic Efficiency, and DeepET\_Topt; AUROC for Cloning CLF and Thermostability; and Accuracy for Material Production and Antibiotic Resistance.

\textbf{(Impact)}  Predicting factors that influence expression levels, stability, and yield can optimize production processes, reducing costs and improving scalability. 

\subsubsection*{Task8: Zero-shot}
\textbf{(Definition \& Metric)} Zero-shot tasks evaluate models' generalization abilities to unseen data without additional training. PFMBench incorporates the ProteinGym dataset \cite{notin2023proteingym}, which assesses the robustness and adaptability of models in predicting mutation effects across diverse proteins. Spearman correlation is the primary metric.

\textbf{(Impact)}  Zero-shot learning is crucial for evaluating models' generalization capabilities, reflecting real-world scenarios where labeled data is scarce or unavailable.

\subsection{More Results}
Table \ref{tab:full_model_performance_adapter} summarizes core model performance across 28 tasks using 6-layer transformer adapters. Sequence-only models performed similarly to ESM2, with no model significantly exceeding the baseline. ProTrek, with contrastive pretraining, achieved the best performance, though potential label leakage from overlapping functional annotation data remains a concern for function-aware models.

\begin{table}[h]
    \caption{Adapter tuning performance of core models on core tasks. }
    \label{tab:full_model_performance_adapter}
    \resizebox{1.0 \columnwidth}{!}{
        \begin{tabular}{lllllllllllll} \toprule
        Model         & ESM-2   & VenusPLM & ESM-C   & ProtGPT2 & PGLM    & ProtT5  & DPLM    & SaProt  & ProstT5 & ProtST  & ESM3    & ProTrek \\ \midrule
        PDBbind       & 0.14677 & 0.16536  & 0.14692 & 0.13503  & 0.16877 & 0.20105 & 0.13659 & 0.15549 & 0.18344 & 0.19514 & 0.15572 & 0.17322 \\
        BindingDB     & 0.13692 & 0.16834  & 0.20716 & 0.17169  & 0.16884 & 0.19730 & 0.17408 & 0.16557 & 0.16642 & 0.18886 & 0.22519 & 0.19230 \\
        M. I. Bin.    & 0.71170 & 0.70195  & 0.70195 & 0.71170  & 0.74513 & 0.72145 & 0.70056 & 0.76323 & 0.72145 & 0.51532 & 0.70334 & 0.80362 \\
        Stability     & 0.32112 & 0.33907  & 0.29976 & 0.14803  & 0.33127 & 0.18638 & 0.29440 & 0.24804 & 0.13032 & 0.06623 & 0.15650 & 0.04924 \\
        Anti. Res     & 0.63422 & 0.64602  & 0.67257 & 0.68437  & 0.67257 & 0.68732 & 0.68732 & 0.65782 & 0.69027 & 0.63422 & 0.58407 & 0.59292 \\
        TCR P. Aff.   & 0.93190 & 0.93784  & 0.93378 & 0.94002  & 0.94542 & 0.93983 & 0.92470 & 0.89967 & 0.93078 & 0.91649 & 0.86510 & 0.90497 \\
        Contact       & 0.71755 & 0.58946  & 0.72026 & 0.07141  & 0.63453 & 0.79012 & 0.71687 & 0.83507 & 0.82642 & 0.52120 & 0.76616 & 0.73618 \\
        Sec. Str.     & 0.76375 & 0.71637  & 0.76777 & 0.49371  & 0.72842 & 0.77978 & 0.75695 & 0.82389 & 0.81397 & 0.68468 & 0.81264 & 0.77363 \\
        FLIP\_AAV     & 0.93848 & 0.92354  & 0.93936 & 0.33732  & 0.87888 & 0.93825 & 0.94491 & 0.94822 & 0.93977 & 0.92250 & 0.92514 & 0.93999 \\
        Fluo.         & 0.68116 & 0.66353  & 0.65043 & 0.61042  & 0.66926 & 0.67662 & 0.67930 & 0.69642 & 0.68020 & 0.56488 & 0.66469 & 0.66987 \\
        Mat. Pro.     & 0.81189 & 0.82018  & 0.81009 & 0.76757  & 0.79495 & 0.80072 & 0.80144 & 0.81081 & 0.81622 & 0.69261 & 0.77514 & 0.81477 \\
        FLIP\_GB1     & 0.95306 & 0.94869  & 0.95772 & 0.86281  & 0.91945 & 0.95217 & 0.92162 & 0.95133 & 0.95408 & 0.82742 & 0.88144 & 0.94049 \\
        Fold Cls      & 0.77546 & 0.75460  & 0.73067 & 0.64724  & 0.77055 & 0.82761 & 0.79448 & 0.80552 & 0.82761 & 0.72577 & 0.72515 & 0.80613 \\
        GO BP         & 0.54411 & 0.54212  & 0.51338 & 0.48536  & 0.52669 & 0.55179 & 0.55989 & 0.53964 & 0.56237 & 0.53352 & 0.41313 & 0.61936 \\
        EC            & 0.73578 & 0.75194  & 0.71694 & 0.69687  & 0.74659 & 0.76201 & 0.75521 & 0.75144 & 0.76829 & 0.71761 & 0.64830 & 0.76408 \\
        GO MF         & 0.64062 & 0.66136  & 0.60517 & 0.58921  & 0.64860 & 0.65762 & 0.66604 & 0.65340 & 0.68076 & 0.62999 & 0.54672 & 0.71195 \\
        GO CC         & 0.61448 & 0.62054  & 0.61501 & 0.58498  & 0.61593 & 0.60873 & 0.62185 & 0.62047 & 0.60785 & 0.63078 & 0.52218 & 0.70202 \\
        DL Bin.       & 0.90619 & 0.91855  & 0.90482 & 0.90117  & 0.91495 & 0.90736 & 0.93305 & 0.92042 & 0.91657 & 0.94016 & 0.90032 & 0.94336 \\
        Sor. Sig.     & 0.87027 & 0.80974  & 0.85391 & 0.77861  & 0.81180 & 0.79012 & 0.83804 & 0.81408 & 0.82789 & 0.87278 & 0.79688 & 0.86161 \\
        DL Multi      & 0.75899 & 0.73502  & 0.76165 & 0.68442  & 0.72437 & 0.69907 & 0.78029 & 0.69241 & 0.73236 & 0.76698 & 0.62051 & 0.80826 \\
        DL2 Multi     & 0.76191 & 0.73814  & 0.75395 & 0.70341  & 0.74772 & 0.72624 & 0.75759 & 0.74006 & 0.73190 & 0.74886 & 0.65853 & 0.83944 \\
        Pep. H/M Aff. & 0.96347 & 0.96616  & 0.96046 & 0.90498  & 0.96638 & 0.95677 & 0.96310 & 0.94768 & 0.95392 & 0.94323 & 0.93000 & 0.94650 \\
        Hum. PPI      & 0.85095 & 0.82147  & 0.83961 & 0.79784  & 0.87692 & 0.81359 & 0.85760 & 0.85100 & 0.79113 & 0.80034 & 0.72483 & 0.84690 \\
        Clo. CLF      & 0.80586 & 0.83172  & 0.81033 & 0.77730  & 0.83638 & 0.78485 & 0.81247 & 0.81206 & 0.79853 & 0.80714 & 0.77391 & 0.82612 \\
        Thermo.       & 0.95036 & 0.91701  & 0.94953 & 0.91401  & 0.94224 & 0.92826 & 0.93949 & 0.96930 & 0.91747 & 0.94393 & 0.87837 & 0.93172 \\
        DeepSol       & 0.84494 & 0.82775  & 0.84171 & 0.78883  & 0.82160 & 0.78741 & 0.82841 & 0.84364 & 0.81937 & 0.81951 & 0.78106 & 0.83427 \\
        DeepSoluE     & 0.77630 & 0.74926  & 0.76009 & 0.68645  & 0.75549 & 0.72004 & 0.74118 & 0.75492 & 0.74905 & 0.72849 & 0.67909 & 0.73090 \\
        ProtSolM      & 0.85874 & 0.84107  & 0.85452 & 0.79735  & 0.84894 & 0.80456 & 0.84847 & 0.85718 & 0.84728 & 0.79923 & 0.80773 & 0.83168 \\ \bottomrule
        \end{tabular}}
\end{table}

\begin{figure}[h]
    \centering
    \includegraphics[width=4.8in]{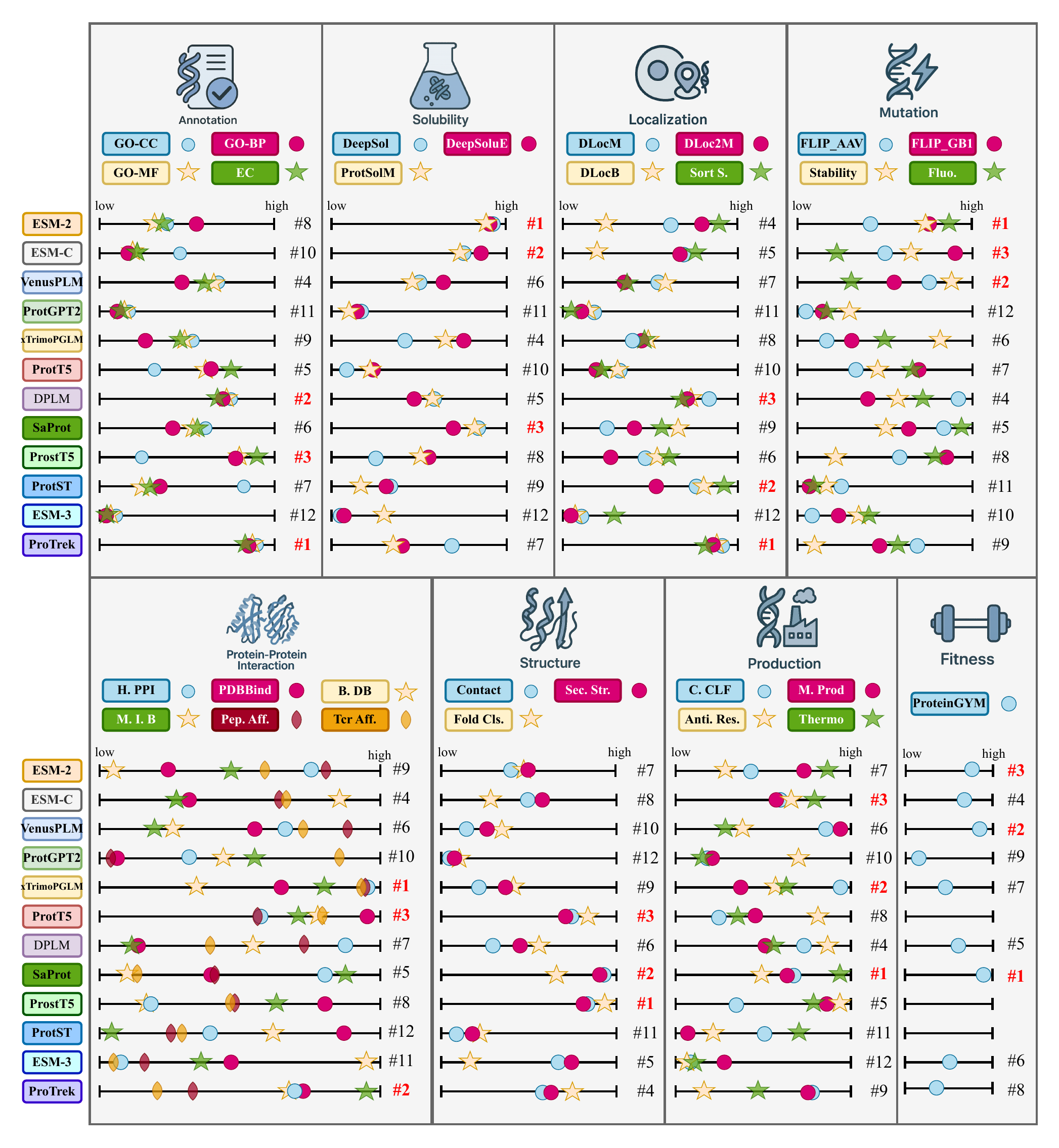}
    \caption{Model rank on tasks.} 
    \label{fig:rank}
\end{figure}

The detailed model rankings across different tasks are shown in Fig.~\ref{fig:rank}, with tasks grouped by category. Different models excel at different types of tasks, such as ProTrek for annotation, ESM2 for solubility, and PGLM for interaction. The zero-shot results do not correlate with the supervised tuning results.

\vspace{-6mm}
\subsection{Mutual Information}
\label{appendix:MI}
\paragraph{Mutual Information Difference Metric.} For a set of MSA sequences $\{ x^{(0)}, x^{(1)}, x^{(2)}, x^{(3)}, \cdots\}$, we compute the mutual information (MI) ~\cite{tschannenmutual,poole2019variational,mcallester2020formal} between the target sequence $x^{(0)}$ and a query sequence $x^{(i)}$. When the two MSA sequences differ in length, the mutual information is computed only over their aligned and overlapping regions. The mutual information is defined as:
\[
I(x^{(i)}; x^{(0)}) = \sum_{k \in \gI} \log \frac{p(x^{(i)}_k \mid x_{/k}^{(0)})}{p(x^{(i)}_k)},
\]
where $\gI$ represents the set of mask indices, $p(x^{(i)}_k \mid x_{/k}^{(0)})$ denotes the conditional probability of the $k$-th token in $x^{(i)}$ predicted by a PLM given the context of $x_{/k}^{(0)}$, $x_{/k}^{(0)}$ indicates that the $k$-th residue is masked, and $p(x^{(i)}_k)$ refers to the marginal probability when the input is fully masked. We use a PLM to estimate $p_{\theta}(x^{(i)}_k \mid x_{/k}^{(0)})$ and compute the MI difference between different PLMs. Taking ESM2-35M as the base model $\theta_0$, the MI difference for a new model $\theta_1$ is defined as:
\[
I(x^{(i)}; x^{(0)}, \theta_1) - I(x^{(i)}; x^{(0)}, \theta_0) 
= \sum_{k \in \gI}\log \frac{p_{\theta_1}(x^{(i)}_k \mid x_{/k}^{(0)})}{p_{\theta_0}(x^{(i)}_k \mid x_{/k}^{(0)})}.
\]

\end{document}